\documentclass[12pt,showpacs,aps,nofootinbib,floatfix]{revtex4}
\usepackage{epsfig}
\usepackage{graphicx}

\newcommand{\be}{\begin{eqnarray}}
\newcommand{\ee}{\end{eqnarray}}

\begin{document}

\title{Dynamics and Stability of Chiral Fluid}
\author{Igor N. Mishustin$^{a,b}$, Tomoi Koide$^{c,d}$, Gabriel S. Denicol$^{c,e}$ and Giorgio Torrieri$^{a,f}$}

\affiliation{ $^a$Frankfurt Institute for Advanced Studies, J.W. Goethe University, Ruth-Moufang Stra{\ss }e 1, 60438 Frankfurt am Main, Germany\\
$^b$National Research Centre Kurchatov Institute, 123182 Moscow, Russia\\ $^c$Institute for Theoretical Physics, J.W. Goethe University, Max von
Laue-Stra{\ss }e 1, 60438 Frankfurt am Main, Germany\\ $^d$Instituto de Fisica, Universidade Federal do Rio de Janeiro, C.P. 68528, 21941-972 Rio de
Janeiro, Brazil\\ $^e$Department of Physics, McGill University, 3600 University Street, Montreal, Quebec, H3A 2T8, Canada\\$^f$Pupin Physics Laboratory, Columbia University, 538 West 120th Street NY 10027, USA }

\begin{abstract}
\noindent {\bf Abstract:} Starting from the linear sigma model with constituent quarks we derive the
chiral fluid dynamics where hydrodynamic equations for the quark fluid are
coupled to the equation of motion for the order-parameter field. In a static system at thermal equilibrium this model leads to a chiral phase transition which, depending on the choice of the
quark-meson coupling constant $g$, could be a crossover or a first order one. We investigate the stability of the chiral fluid in the static and
expanding backgrounds by considering the evolution of perturbations with respect to the mean-field solution. In the static background the spectrum of
plane-wave perturbations consists of two branches, one corresponding to the sound waves and another to the $\sigma$-meson excitations. For large $g$
these two branches "cross" and the excitation spectrum acquires exponentially growing modes. The stability analysis is also done for the Bjorken-like
background solution by explicitly solving the time-dependent differential equation for perturbations in the $ \eta$ space. In this case the growth
rate of unstable modes is significantly reduced.
\end{abstract}

\pacs{25.75.-q, 47.75.+f, 11.30.Qc, 24.60.Ky}

\maketitle

\hbadness=10000

\topmargin -0.8cm\oddsidemargin = -0.7cm\evensidemargin = -0.7cm


\section{Introduction}

Creation of new forms of strongly interacting matter, in particular, the observation of a deconfined and chirally restored state, is the main goal of present and future experiments with relativistic heavy-ion beams \cite{fair,rhicstar,rhicphe,nica,shine}. Significant progress in understanding the
dynamics of relativistic heavy-ion collisions and the properties of dense matter produced in such collisions has been achieved in experiments at
Relativistic Heavy Ion Collider (RHIC) \cite{whiteBRAHMS,whitePHENIX,whitePHOBOS,whiteSTAR}. Recently the Large Hadron Collider (LHC) has started a new era in high-energy nuclear physics
by colliding p+p and Au+Au beams at much higher energies than in previous experiments, and new interesting results have been obtained already, see, e.g. \cite{Mueller} and references therein.
The investigations of the phase diagram of strongly interacting matter in the temperature-chemical potential plane and, in particular, searching for manifestations of the QCD-based phase transitions remains in the focus of theoretical and experimental studies.

The signatures of such phase transitions have been studied mostly based on equilibrium concepts. However, the process of a relativistic heavy-ion
collision at RHIC and LHC energies is very fast and one should expect that the phase transition may be strongly affected by the dynamics. Thus, it is
important to study the phase transition dynamics in a time-dependent background. Previously non-equilibrium effects associated with the
chiral/deconfinement phase transition have been studied within several macroscopic approaches
\cite{WR,CM,Rand1,mish1,Scavenius2,mish2,Scaven2,Rand2,Rand3,Tomasik,Fraga,kunihiro,skokov,Jan}.

Generally, in order to investigate non-equilibrium effects, one should solve a quantum many-body problem by using, e.g. Kadanoff-Baym equation. But
applying such formalism to relativistic heavy-ion collisions is very complicated, see, e.g. Ref.\ \cite{PHSD}. Fortunately, there is still hope that
the collective behavior of the hot matter created in heavy-ion collisions can be described by a coarse-grained macroscopic theory, such as
hydrodynamics. Indeed, some aspects of relativistic heavy-ion collisions, such as the collective flow of the produced matter, are well described in the
framework of hydrodynamic models \cite {heinz,shuryak,romatschke,huovinen,chojnacki,hirano,Sat1,Schenke,Sat2}. However, in order to describe a phase
transition in a time-dependent background, the usual hydrodynamic model should be modified by explicitly considering the dynamics of the order
parameter. This becomes necessary when the time scale associated with the order parameter relaxation becomes of the same order or longer than the
characteristic time associated with the hydrodynamic variables. In the opposite limit, the effect of a phase transition can be taken into account
through the EoS, as is usually done in macrophysics(see e.g. Ref.\ \cite{Sat2}).

In this paper we use a modified hydrodynamic theory, namely the Chiral Fluid Dynamics (CDF), in which the fluid evolution is coupled to the dynamics of
the chiral order parameter. This model was firstly proposed in Ref.\ \cite{mish4} and further developed in several works \cite
{paech,davesne,pratt,kodama}. Recently it was generalized by including fluctuations of the order parameter and dissipative terms
\cite{Nahrgang1,Nahrgang2,Herold}. It can be derived from the linear sigma model by assuming that microscopic and macroscopic degrees of freedom are
clearly separated. Then the coarse-grained macroscopic dynamics is described only by a reduced number of variables, which are called the gross
variables \cite{onuki,koide1,koide2}.

As was first pointed out by van Hove \cite{hove}, the relaxation time of the order parameter increases near the second order phase transition and it
diverges at the critical point. This phenomenon is known as "critical slowing down" \cite{L&L}. Its importance for modeling phase transitions in
dynamical environments was demonstrated in ref. \cite{slowing}. In this situation non-equilibrium effects need to be considered explicitly even in an
ideal fluid. This can be done by choosing the order parameter as another gross variable.

In the region of the phase diagram where the deconfinement/chiral phase transition is of first order, an extra time-scale appears, which is associated
with the nucleation process \cite{kapusta}. Only when this time scale is short with respect to the hydrodynamic time scale, we can assume a two-phase
equilibrium and the equation of state given by the Maxwell construction. However, if it is not the case, the system will pass through a
metastable region of the equation of state and can even reach the point of spinodal instability \cite{WR,CM,Rand1,mish1,Jan,Herold}. In order to
investigate this possibility, we study the stability of fluctuations around a hydrostatic state and a Bjorken-type expanding state. By comparing the
results, we will be able to identify those features of the phase transition which are affected by the fast dynamics.

The paper is organized as follows. In Section II we derive basic equations of CFD from the linear $\sigma $-model with constituent quarks. In Section
III we discuss the predictions of this model for the equilibrated system undergoing a chiral phase transition. Then in Section IV we calculate the
excitation spectrum of the system by introducing fluctuations around the static background solution. In Section V we study the time evolution of
fluctuations in the Bjorken-like expanding background. Our concluding remarks are presented in Section VI.

\section{Derivation of chiral fluid dynamics}

As the low-energy effective theory of QCD, we adopt the linear sigma model with constituent quarks \cite{Gell} whose qualitative features (chiral
symmetry, universality class, phase transition structure) are thought to coincide with QCD \cite{LeeWick,PW}. More recently the thermodynamics of this
model was studied on the mean-field level \cite{Scavenius1}, as well as including the field fluctuations \cite{Scadron,bbs,mish3,Kap09}. Following the
previous works \cite{mish4,paech}, we describe the coarse-grained dynamics of the quark degrees of freedom with the hydrodynamic variables, coupled to
the order parameter field $\sigma $ via its equation of motion.

The Lagrangian of the linear $\sigma $ model is
\begin{equation} \label{Lag}
\mathcal{L}=\bar{q}(i\gamma ^{\mu }\partial _{\mu }-g(\sigma +i\gamma _{5} \vec{\tau}\vec{\pi}))q+\frac{1}{2}\left[ (\partial _{\mu }\sigma)^{2}+(\partial _{\mu }\vec{\pi})^{2}\right] -V(\sigma ,\pi ),
\end{equation}
where $q$ is the quark field, $\sigma $ and $\vec{\pi}$ are the chiral fields, $g$ is the quark-meson coupling constant. The \textquotedblleft Mexican
Hat\textquotedblright\ potential $V$ is given by
\begin{equation}
V(\sigma ,\vec{\pi})=\frac{\lambda ^{2}}{4}(\sigma ^{2}+(\vec{\pi} )^{2}-v^{2})^{2}-H\sigma ,  \label{pote}
\end{equation}
where $\lambda $, $v$ and $H$ are the model parameters. They can be calculated by using the physical pion mass $m_{\pi }=138$ MeV, the pion decay
constant $f_{\pi }=93$ MeV and the sigma mass in vacuum $m_{\sigma }=600$ MeV: $H=f_{\pi }m_{\pi }^{2}$, $v^{2}=f_{\pi }^{2}-m_{\pi }^{2}/\lambda ^{2}$
and $\lambda ^{2}=(m_{\sigma }^{2}-m_{\pi }^{2})/(2f_{\pi }^{2})$. This potential provides a mechanism for spontaneous breaking of the chiral symmetry
with non-zero vacuum expectation value of the sigma field, $\sigma _{\mathrm{vac}}=f_{\pi }$.

The coupling constant $g$ is usually chosen so that the constituent quark mass in vacuum, $g\sigma _{\mathrm{vac}}$, is equal to one-third of the
vacuum nucleon mass, that is, $g=3.3$. Then the chiral phase transition at the vanishing chemical potential is shown to be of the crossover type. But
the phase diagram has a critical (end) point at a finite baryon chemical potential, where the order of the phase transition changes to first order
\cite{Scavenius1}. In this paper we limit our consideration to the case of vanishing chemical potential. In this case the effect of the first order
phase transition can be studied by changing the magnitude of the coupling constant $g$, as was proposed in Ref.\ \cite{Scavenius2}. In order to study
the dynamics of a first order phase transition, we shall also consider the case of $g=4.5$, in which the phase transition is of first order.

For the sake of simplicity, the pion field is neglected in the rest of this paper and the sigma field $\sigma $ is considered as a classical field
(condensate). We consider an idealized situation where the quark degrees of freedom have already achieved local thermal equilibrium and can be
approximately described as an ideal fluid, characterized by the energy density $\varepsilon $, the pressure $P$ and the fluid four-velocity $u^{\mu }$.
On the other hand, the sigma field behaves as a classical external field (the chiral order parameter) acting on quarks through the mass term $
M=g\sigma $. Then the energy momentum tensor of the system can be represented as
\begin{equation}
T^{\mu \nu }=T_{\mathrm{fluid}}^{\mu \nu }+T_{\mathrm{field}}^{\mu \nu }. \label{eq2}
\end{equation}
The energy-momentum tensor of an ideal fluid is generally represented as
\begin{equation}
T_{\mathrm{fluid}}^{\mu \nu }=(\varepsilon +P)u^{\mu }u^{\nu }-Pg^{\mu \nu }, \label{Tmunufluid}
\end{equation}
where $\varepsilon $ and $P$ are the proper scalar densities, which coincide with the energy density and pressure of the fluid in the rest frame.

These quantities are calculated via the thermodynamic potential of the quark sector calculated in the mean-field approximation,
\begin{equation}
\Omega (M)=-\nu _{q}T\int \frac{d^{3}\mathbf{p}}{(2\pi )^{3}}\left\{ \ln { \left[ 1+\exp {\left( \frac{\mu -E_{p}}{T}\right) }\right] }+\ln {\left[1+\exp {\left( \frac{-\mu -E_{p}}{T}\right) }\right] }\right\} ~, \label{Om_M}
\end{equation}
where $T$, $\mu $ and $E_{p}$ are temperature, quark chemical potential and particle energy $\sqrt{\mathbf{p}^{2}+M^{2}}$ with $M$ being the
constituent quark mass, respectively. The degeneracy factor $\nu _{q}$ is given by $ 2N_{c}N_{f}$=12 with $N_{c}$ and $N_{f}$ being the number of
colors and flavors, respectively. With this thermodynamic potential, the pressure and the energy density are, respectively, given by
\begin{equation}
P(M)=-\Omega (M)=\frac{\nu _{q}}{3}\int \frac{d^{3}\mathbf{p}}{(2\pi )^{3}} \frac{\mathbf{p}^{2}}{E_{p}}\left[ f(E_{p}-\mu )+f(E_{p}+\mu )\right] ~,
\label{press}
\end{equation}
\begin{equation}
\varepsilon (M)=\Omega -T\frac{\partial \Omega }{\partial T}=\nu _{q}\int
\frac{d^{3}\mathbf{p}}{(2\pi )^{3}}
E_{p}\left[ f(E_{p}-\mu )+f(E_{p}+\mu )\right]
~,  \label{enerden}
\end{equation}
where we have introduced the Fermi distribution function, $f(E)=\left( e^{\frac{E}{T}}+1\right) ^{-1}$. As was already mentioned above, in the
following calculations we consider only the case of $\mu =0$. Strictly speaking, the above expressions are valid only when $M$ is a space-time
independent constant determined by the minimum of the total thermodynamic potential,
\begin{equation}
\Omega _{\mathrm{tot}}(\sigma ,\mathbf{\pi })=\Omega (M)+V(\sigma ,\vec{\pi} )~.  \label{omtot}
\end{equation}
However, in our exploratory study below, we assume that $M=g\sigma (x)$ even when the $\sigma (x)$ is varying in space and time according to the
equation of motion obtained from the Lagrangian,
\begin{equation}
\partial ^{2}\sigma +\lambda (\sigma ^{2}-\sigma _{0}^{2})\sigma -H=-g\bar{q}
q.  \label{eq_sigma}
\end{equation}
Further on, we replace the term $\bar{q}q$ in the r.h.s. by its thermal expectation value, that is, the quark scalar density
\begin{equation}
\langle \bar{q}q\rangle \equiv \rho _{s}(x)=2\nu _{q}\int \frac{d^{3}\mathbf{p}}{(2\pi )^{3}}\frac{M}{E_{p}}f(E_{p})~.  \label{qsource}
\end{equation}
By combining Eqs.(\ref{press}) and (\ref{enerden}), one can easily check the relation:
$\varepsilon -3P=M\rho _{s}$.

After separating the quark contribution, the energy-momentum tensor of the $ \sigma $ field is obtained from the meson part of Lagrangian (\ref{Lag})
\footnote{ It is however known that this definition of the energy-momentum tensor involves serious problems \cite{callan,huang}.}
\begin{equation}
T_{\mathrm{field}}^{\mu \nu }=(\partial ^{\mu }\sigma )(\partial ^{\nu }\sigma )-g^{\mu \nu }\left[ \frac{1}{2}(\partial _{\mu }\sigma )^{2}-V(\sigma
,0)\right] ~.  \label{Tmunufield}
\end{equation}
Now the continuity equation for the total energy-momentum tensor (\ref{eq2} ), $\partial _{\nu }T^{\mu \nu }=0$, can be written as
\begin{equation}
\partial _{\nu }T_{\mathrm{fluid}}^{\mu \nu }=-\partial _{\nu }T_{\mathrm{
field}}^{\mu \nu }\equiv S^{\mu }~,  \label{fluideq}
\end{equation}
where the source term is given by
\begin{equation}
S^{\mu }=-\left[ \partial ^{2}\sigma +\lambda ^{2}\left( \sigma ^{2}-\sigma _{0}^{2}\right) \sigma -H\right] \partial ^{\mu }\sigma =g\rho _{s}\partial
^{\mu }\sigma .  \label{eq_emtensor}
\end{equation}
This term gives rise to the dynamical coupling between ideal fluid dynamics and the chiral order parameter.

Note that dissipative effects, such as the viscosity of the fluid and damping of the field fluctuations, are neglected in the present approach. On the
other hand, our approach includes the derivative terms of the sigma field and hence the energy associated with the spatial and temporal inhomogeneity
of the system. Inhomogeneity of thermodynamic quantities are usually ignored in hydrodynamical calculations, although they may play an essential role
in nucleus-nucleus collisions, especially in the dynamical phase transitions proceeding via the spinodal decomposition or the nucleation.

\section{Mean-field results in thermal equilibrium}

In this Section we discuss thermodynamic properties of the chiral fluid. The thermodynamics of the corresponding linear sigma model was studied on the
mean-field level \cite{Scavenius1}, as well as including the field fluctuations \cite{Scadron,bbs,mish3,Kap09}. Below we only summarize the features
which are important for our further analysis.

\subsection{Effective thermodynamic potential}

\begin{figure}[h]
\includegraphics[scale=0.25]{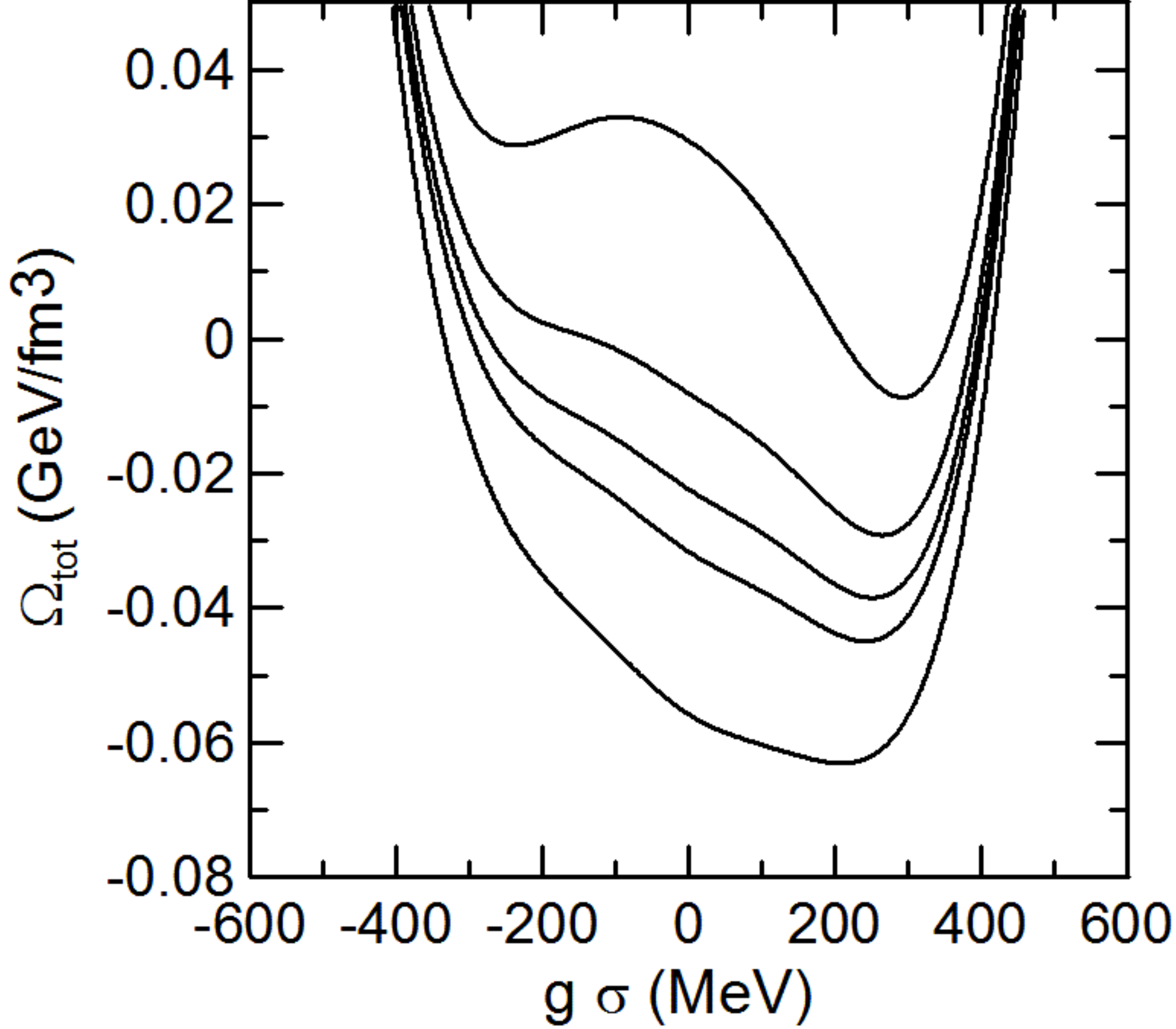} %
\includegraphics[scale=0.25]{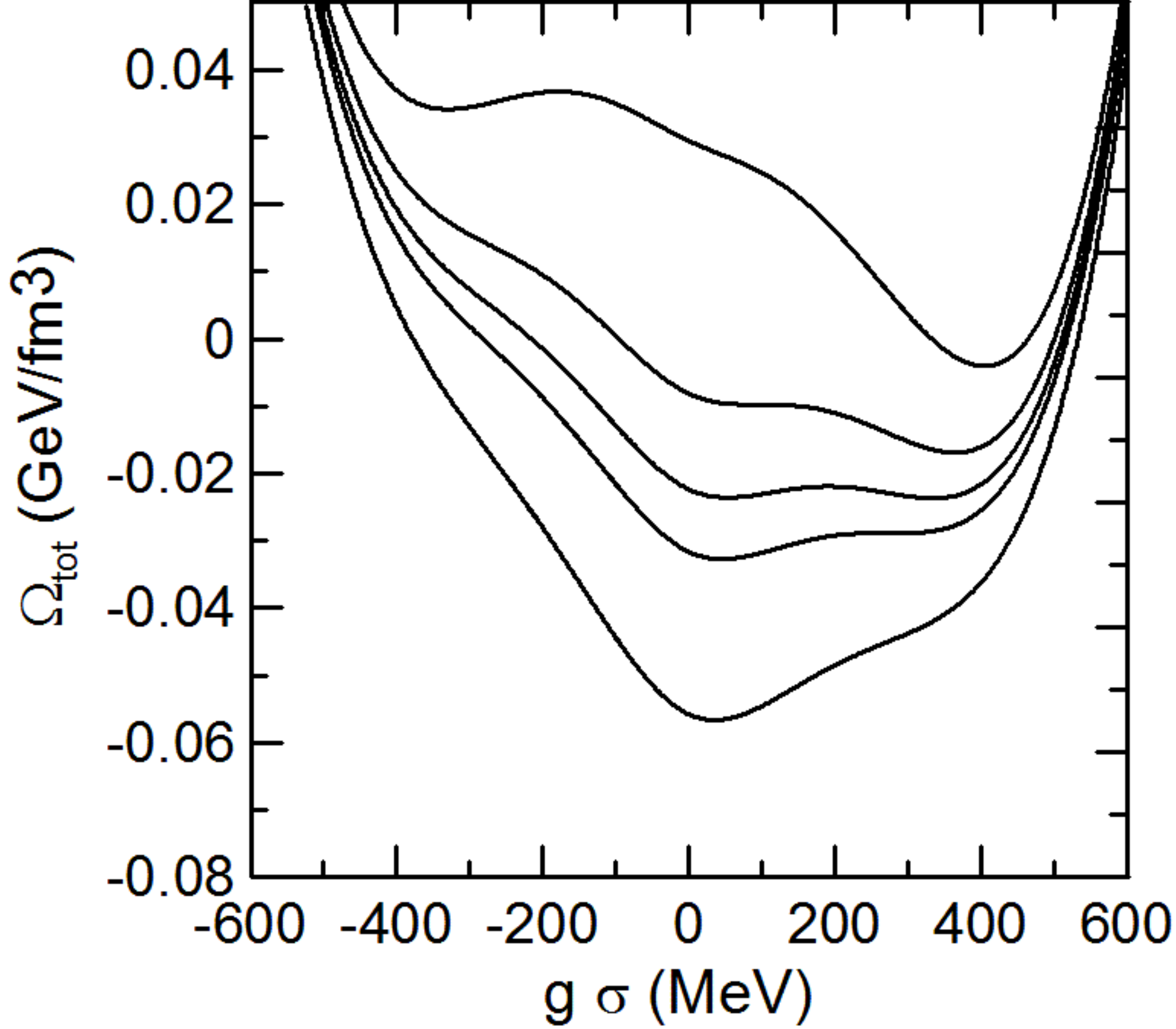}
\caption{The temperature dependence of the total thermodynamic potential for $g=3.3$ (left) and $g=4.5$ (right). The lines from top to bottom
correspond to $T=100$, $122.7$, $128.5$, $132.1$ and $140$ MeV.} \label{TDPfig}
\end{figure}

The total thermodynamic potential $\Omega _{\mathrm{tot}}$, defined by Eqs.\ ( \ref{Om_M}) and (\ref{omtot}), is shown in Fig.\ \ref{TDPfig} as a
function of the chiral order parameter $\sigma $. The calculations are made for several values of $T$ fixing $\mu =0$. Two panels show results for two
different choices of the coupling constant $g$. For $g=3.3$, the sequence of curves is typical for the crossover type of phase transition. The
potential has only one minimum for $\sigma >0$ which moves gradually from values close to the vacuum value of the sigma field at low $T$ to almost zero
at high $T$. \footnote{ The "minimum" at $\sigma <0$ in fact becomes a saddle point, when the pion field is included in the consideration.} In
contrast, for $g=4.5$, the potential has two local minima in the temperature interval between $T_{ \mathrm{D}}=122.7$ MeV and $T_{\mathrm{B}}=132.1$
MeV. They represent two competing phases, corresponding to broken (large $\sigma $) and partially restored (small $\sigma $) chiral symmetry. This is
the characteristic behavior of a first order phase transition. The transition temperature is determined by the condition that two minima have equal
height, $T_{\mathrm{C}}=128.5$ MeV in the considered case. A similar behavior of the thermodynamic potential was found in Ref.\ \cite{Scavenius1} for
$g=3.3$ but for finite values of the quark chemical potential, $\mu >207$ MeV.

\begin{figure}[tbp]
\includegraphics[scale=0.3]{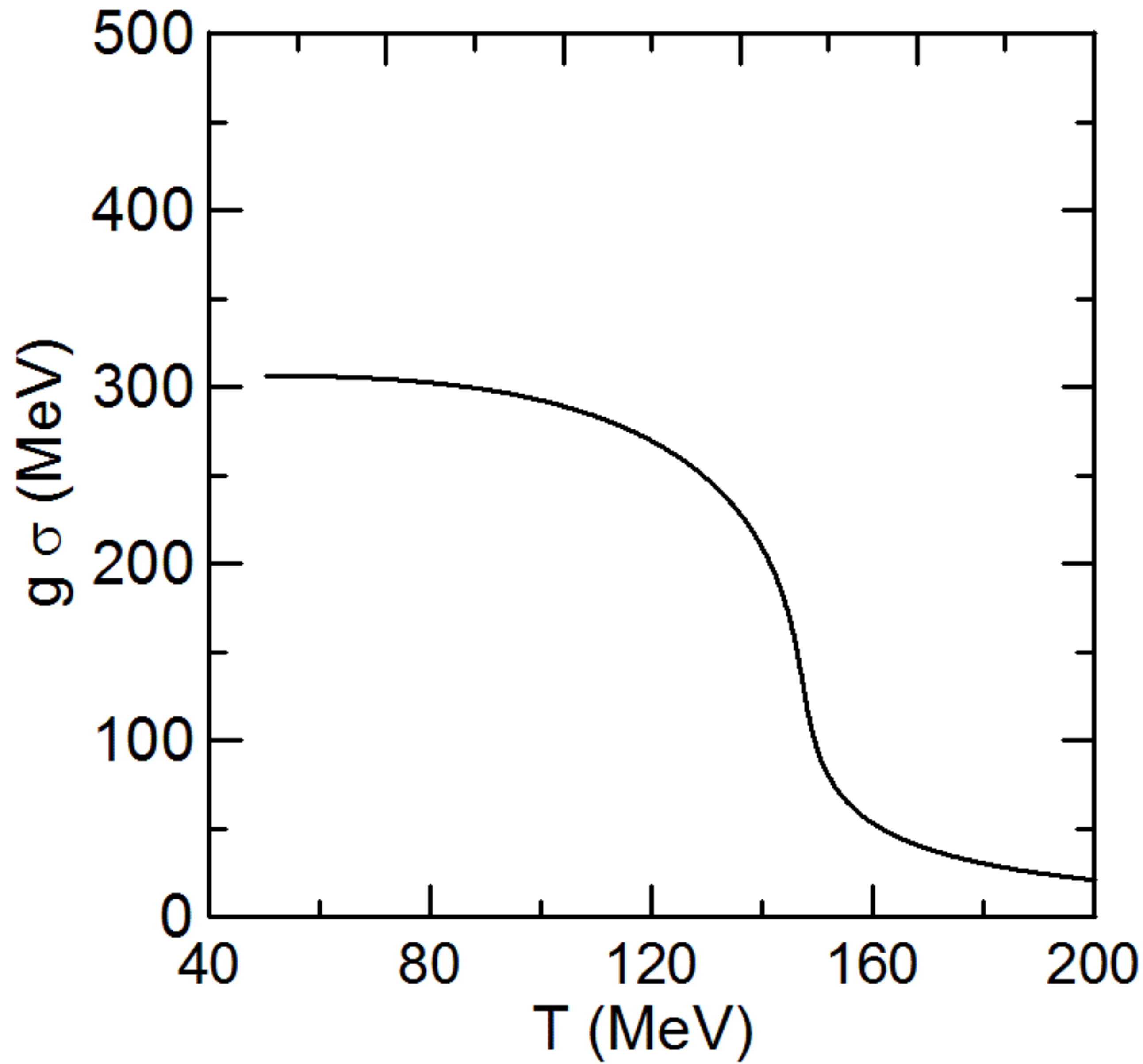}
\includegraphics[scale=0.3]{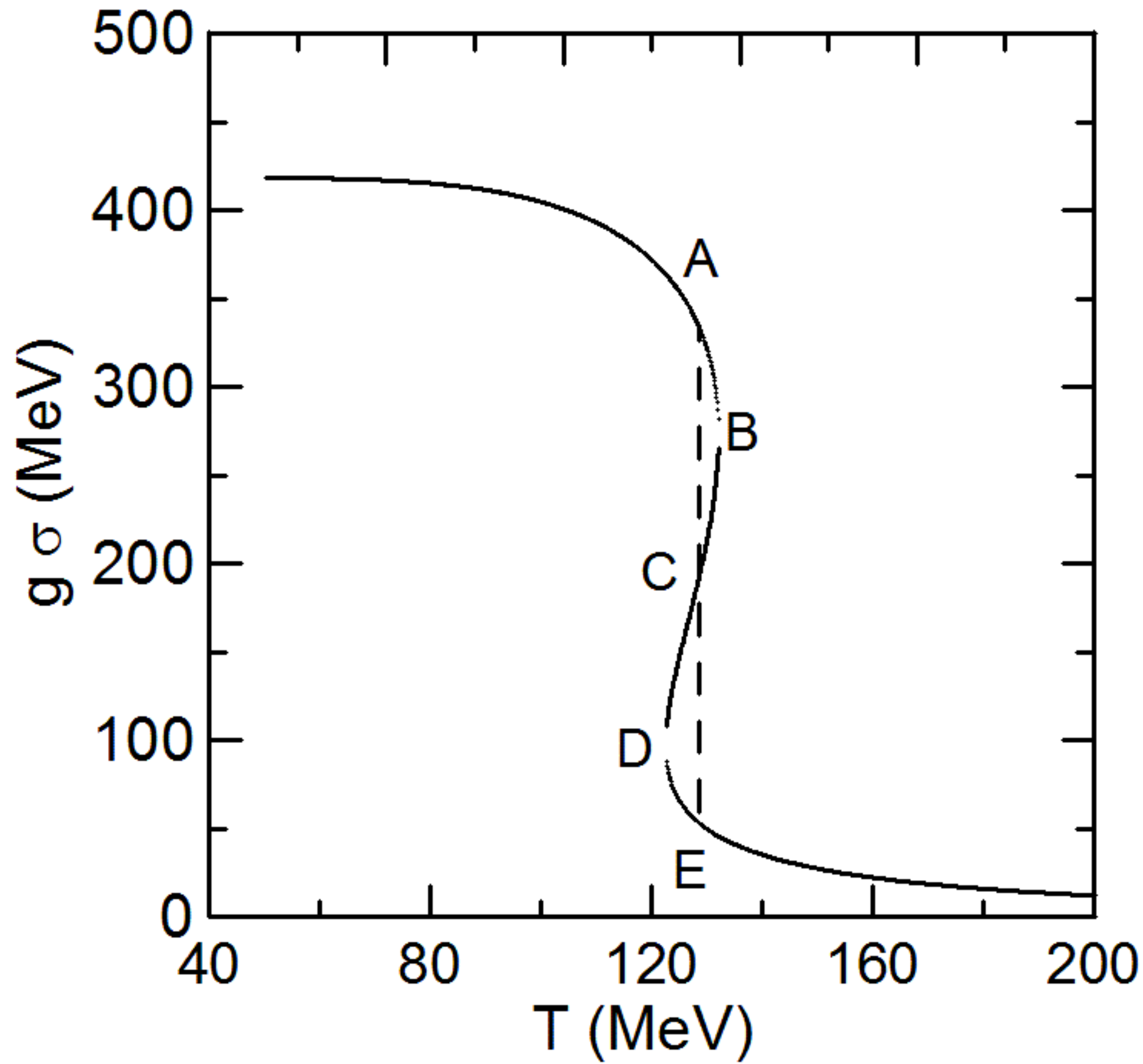}
\caption{The temperature dependence of $g\protect\sigma_0$ for $g=3.3$ (left panel) and $g=4.5$ (right panel). The dashed line AE shows the Maxwell
construction for the first order phase transition. States between points B and D correspond to local maxima of the thermodynamic potential. }
\label{gapfig}
\end{figure}

The equilibrium condensate $\sigma _{0}$ corresponds to the minimum of the total thermodynamic potential $\Omega _{\mathrm{tot}}(\sigma )$, i.e., it is
found from the equation
\begin{equation}
\frac{\partial \Omega _{\mathrm{tot}}}{\partial \sigma }=\lambda ^{2}\sigma (\sigma ^{2}-v^{2})-H+g\rho _{s}(\sigma ,T)=0~.  \label{gapeq}
\end{equation}
This condition is equivalent to the equation of motion for the sigma field, Eq.\ (\ref{eq_sigma}), in the static and uniform background. It is obvious
that Eq.\ (\ref{gapeq}) alone cannot guarantee that the solution corresponds to a minimum. One should also check the second derivative of the effective
potential at $\sigma =\sigma _{0}$, which is usually associated with the effective sigma mass square,
\begin{equation}
m_{\sigma }^{2}(\sigma _{0},T)=\left( \frac{\partial ^{2}\Omega _{\mathrm{tot }}}{\partial \sigma ^{2}}\right) _{\sigma _{0},T}=\lambda ^{2}\left(
3\sigma _{0}^{2}-v^{2}\right) +g\left( \frac{\partial \rho _{s}}{\partial \sigma } \right) _{\sigma _{0},T}~.  \label{sigmass}
\end{equation}
As we will see below, in general Eq.\ (\ref{gapeq}) has all types of solutions corresponding to minima ($m_{\sigma }^{2}>0$), maxima ($m_{\sigma
}^{2}<0$) and inflection points of the thermodynamic potential. For a specific value of the coupling constant, i.e., $g=3.63$, the potential has a very
flat minimum with a vanishing effective mass, $m_{\sigma }^{2}=0$, which corresponds to the critical point of the second order phase transition
\cite{Nahrgang1}.

In Fig.\ \ref{gapfig}, the temperature dependence of the chiral order parameter $\sigma (T)$ is shown for $g=3.3$ (left panel) and $g=4.5$ (right panel). For $g=3.3$, the condensate is decreasing gradually with temperature. On the other hand, for $g=4.5$, it is a multi-valued function of $T$ at
temperatures between $T_{\mathrm{D}}=122.7$ MeV (point D) and $T_{\mathrm{B}}=132.1$ MeV (point B). In this interval, the thermodynamic potential has
two local minima (above point B and below point D) and one maximum (on segment BCD), as clearly seen in Fig.\ \ref{TDPfig}. In the idealized
equilibrium situation, the first order phase transition may start when the two local minima have equal heights (points A and E) and the transition
follows the dashed line AE at temperature $T_{\mathrm{C}}=128.5$ MeV (Maxwell construction). However, the formation of a new phase is possible only via
penetration through the potential barrier between these two minima, which requires a finite time. Moreover, a certain degree of supercooling is needed
to compensate for the interface energy associated with the creation of an island of a new phase surrounded by the old one \cite {kapusta}. In reality,
this nucleation mechanism works only when the evolution of the potential in the course of system's expansion is sufficiently slow. Otherwise, after
reaching the transition temperature the system will remain in the metastable state until the barrier between two minima disappears. This happens when
the metastable minimum and the maximum of the potential fuse and produce an inflection point (the points B and D in Fig.\ \ref{gapfig}). Then the phase
transition proceeds via the spinodal decomposition (see below).

\subsection{Equation of state of the chiral fluid}

The equilibrium energy density and pressure of the system in the rest frame of the fluid ($v=0$) can be obtained from the total energy-momentum tensor
( \ref{eq2}) taking into account Eqs.\ (\ref{Tmunufluid}) and (\ref{Tmunufield} )
\begin{eqnarray}
\varepsilon _{\mathrm{tot}} &=&T^{00}=\varepsilon (M)+V(\sigma )+\frac{1}{2}\left[ (\partial _{t}\sigma )^{2}+(\nabla \sigma )^{2}\right] ~, \\
P_{\mathrm{tot}} &=&\frac{1}{3}\sum_{i=1}^{3}T^{ii}=P(M)-V(\sigma )+\frac{1}{2}\left[ (\partial _{t}\sigma )^{2}-(\nabla \sigma )^{2}\right]
\end{eqnarray}
where $\varepsilon (M)$ and $P(M)$ are the energy density and pressure of the fluid defined in Eqs.\ (\ref{enerden}) and (\ref{press}), respectively,
and $V(\sigma )=V(\sigma ,\pi =0)$ is given by Eq.\ (\ref{pote}). In thermodynamic calculations it is usually assumed that the system is static and
uniform, i.e. the space-time derivatives of the order parameter field vanish.\footnote{As already mentioned above, the derivative terms may be
important in case of nuclear collisions where the system is certainly inhomogeneous and varying in time.}
Then the equation of state is
obtained by substituting the equilibrium value of the sigma field which is determined by the gap equation (\ref{gapeq}).

\begin{figure}[tbp]
\includegraphics[scale=0.3]{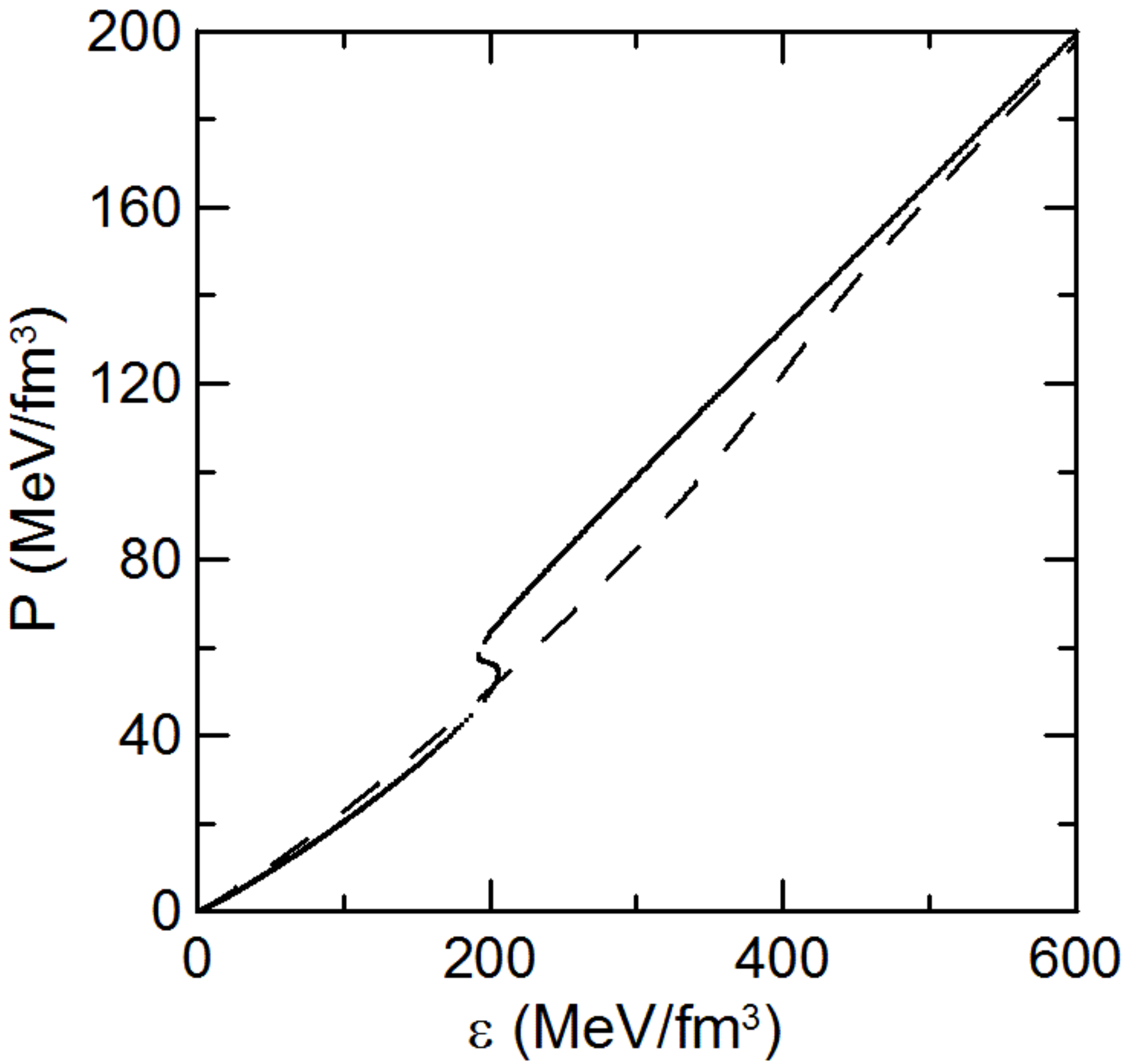} %
\includegraphics[scale=0.3]{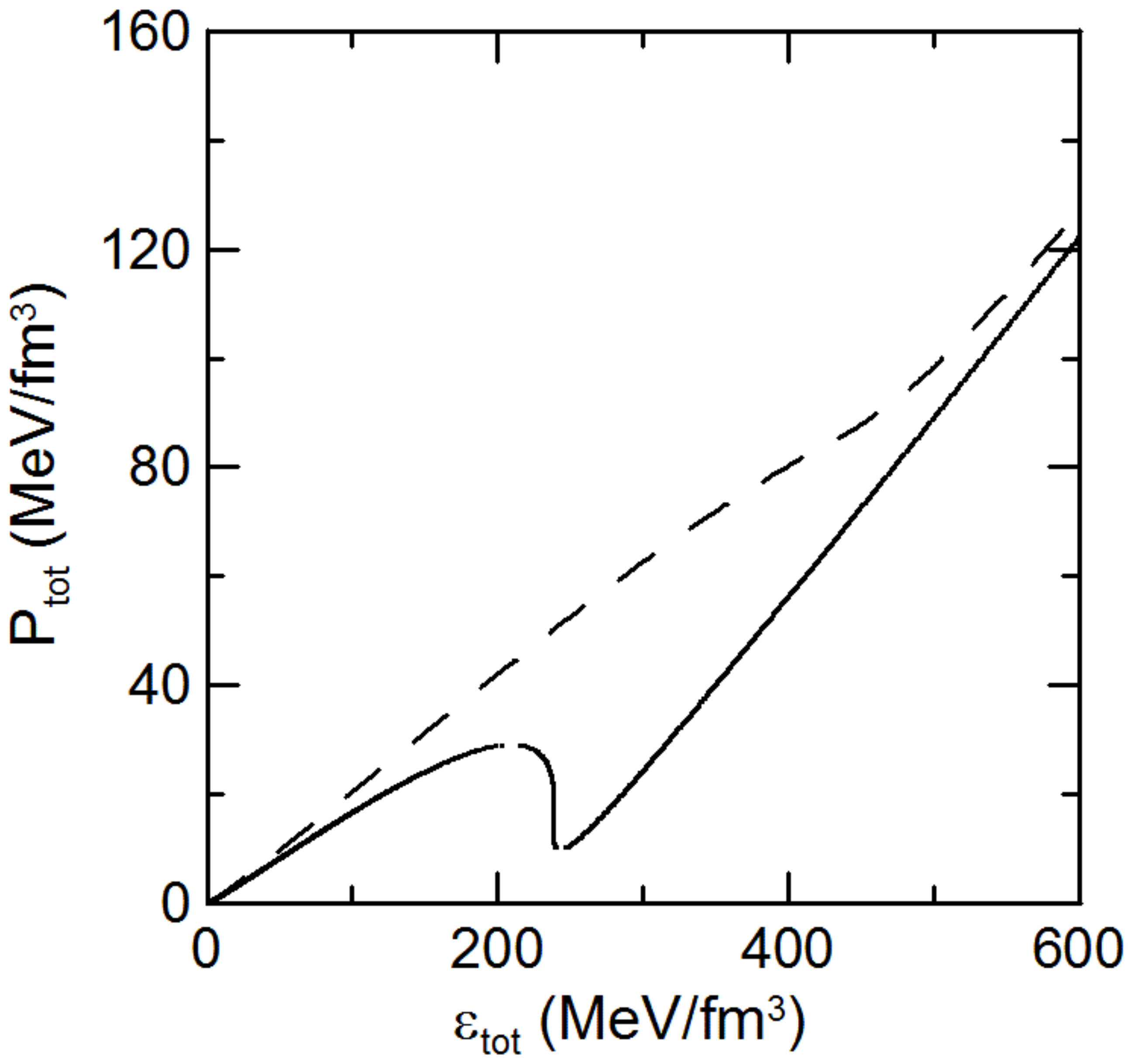}
\caption{The energy density dependence of the pressure of the quark sector (left) and the total pressure (right). The dashed and solid lines represent
$g=3.3$ and $4.5$, respectively. In all cases the curves are shifted so that the energy density and pressure vanish in vacuum. } \label{pefig}
\end{figure}

In Fig.\ \ref{pefig}, the pressure as a function of energy density is shown for the quark fluid alone (left) and for the total system of quark fluid
plus sigma field (right). The dashed and solid lines represent $g=3.3$ and $4.5$, respectively. In the left panel, we see that, for $g=3.3$, the
pressure is a monotonically increasing function of $\varepsilon $ and, in practice, is very close to the equation of state of an ideal gas of massless
particles, i.e., $P=\varepsilon /3$, . On the other hand, for $g=4.5$ the behavior is more complex: the total pressure $P_{\mathrm{tot}}$ becomes a
multi-valued function of the total energy density $\varepsilon _{\mathrm{tot}}$, as is shown by the solid line on the right panel. In other words,
there exists a region in the phase space of thermodynamic variables that has a negative sound velocity square, $c_{s}^{2}=dP_{tot}/d\varepsilon
_{tot}<0$. This is consistent with the mean-field result that a first order phase transition is signaled by spinodal instability ($c_{s}^{2}<0$).

It is interesting that this instability shows up also in the relation between $P$ and $\varepsilon $, characterizing the quark sector alone.
The above argument is not directly applicable to the CFD, because it assumes that only the quantities in the quark sector are thermalized and the pressure there is given by $P$ but not by $P_{tot}$. This is why in Fig. \ref{pefig} (left) we plot $P$ as a function of $\varepsilon$.
As one can see from Fig.\
\ref{pefig} (left), in this case $\varepsilon $ becomes a multi-valued function of $P$, differently from $\varepsilon _{tot}$. Therefore, in this
case too there exists a region where $dP/d\varepsilon <0 $, i.e., the sound velocity square of the quark fluid becomes negative. Thus the instability
associated with the first order phase transition can exist in the chiral fluid even when the order parameter field is out of equilibrium. This point
will be investigated in the following sections in more detail.

\subsection{Generalized thermodynamic relations}

In CFD the chiral order parameter is considered as an independent variable. Thus the thermodynamic relations between $\varepsilon $ and $P$ should be
modified. Noting that the pressure $P$ is a function of $T$ and $\sigma $, we can write
\begin{equation}
dP=sdT-g\rho _{s}d\sigma ~,  \label{Gibbs_Duhem}
\end{equation}
where we have used the relation
\begin{equation}
\left( \frac{\partial P}{\partial \sigma }\right) _{T}=-\left( \frac{
\partial \Omega }{\partial \sigma }\right) _{T}=-g\rho _{s}~,
\end{equation}
which follows from Eqs.\ (\ref{Om_M}) and (\ref{qsource}). We further introduced the entropy density as
\begin{equation}
s=\left( \frac{\partial P}{\partial T}\right) _{\sigma }~.
\end{equation}
Using this definition together with Eqs.\ ({\ref{press}) and (\ref{enerden}), one obtains the standard Gibbs-Durham relation,
\begin{equation}
Ts=\varepsilon +P~.  \label{eq_GD}
\end{equation}
Then the first law of thermodynamics can be written as
\begin{equation}
Tds=d\varepsilon -g\rho _{s}d\sigma ~.  \label{1st_law}
\end{equation}
Such thermodynamic relations have been also discussed in Ref.\ \cite{kodama}. They can be used to derive the following formulas, which will be used
later, }

\begin{equation}
T\left(\frac{\partial s}{\partial T}\right)_{\sigma}=\left(\frac{\partial \varepsilon}{\partial T}\right)_{\sigma},~~ T\left(\frac{\partial s}{
\partial \sigma}\right)_{T}=\left(\frac{\partial \varepsilon}{\partial \sigma
}\right)_{T}-g\rho_s~.
\end{equation}

\section{Spectrum of fluctuations in a static background \label{secstatic}}

In this section, we investigate the stability of the chiral fluid with respect to perturbations of a the static background characterized by the chiral
order parameter $\sigma _{0}$, temperature $T_{0}$ and four-velocity $u_{0}^{\mu }=(1,0,0,0)$. Obviously, when $\sigma _{0}$ is initially chosen on
the slope or at the maximum of the thermodynamic potential, at later times it will roll down to reach the minimum of the potential. Thus, without any
analysis, we know that such a state is unstable. Such situation corresponds to the spinodal decomposition. Thus, in the following, we discuss the
stability only around the local minima of the thermodynamic potential where $\sigma _{0}$ is the solution of the gap equation (\ref{gapeq}).

Let us introduce the plan-wave perturbations of these quantities in the $x$ direction around the hydrostatic state:
\begin{eqnarray}
\delta\sigma(x)&=&\delta \sigma(\omega,k) e^{i\omega t - i k x}~, \\ \delta T(x)&=&\delta T (\omega,k) e^{i\omega t - i k x}~, \\ \delta
u^1(x)&=&\delta u^1(\omega,k) e^{i\omega t - i k x}~.
\end{eqnarray}

Then, the perturbed fluid characteristics can be expressed as
\begin{eqnarray}
\varepsilon (\sigma,T) &=& \varepsilon (\sigma_0, T_0) + \left( \frac{
\partial \varepsilon (\sigma,T)}{\partial T} \right)_0 \delta T(x) + \left(
\frac{\partial \varepsilon (\sigma,T)}{\partial \sigma} \right)_0 \delta \sigma(x)~, \\ P (\sigma,T) &=& P (\sigma_0, T_0) + \left( \frac{\partial P
(\sigma,T)}{\partial T} \right)_0 \delta T(x)+ \left( \frac{\partial P (\sigma,T)}{
\partial \sigma} \right)_0 \delta \sigma(x)~, \\
\rho_s (\sigma,T) &=& \rho_s (\sigma_0, T_0) + \left( \frac{\partial \rho_s (\sigma,T)}{\partial T} \right)_0 \delta T(x)+ \left( \frac{\partial \rho_s
(\sigma,T)}{\partial \sigma} \right)_0 \delta \sigma(x)~,
\end{eqnarray}
where index 0 corresponds to the quantity taken at $\sigma=\sigma_0$ and $T=T_0$.

By linearizing the fluid dynamical equations (\ref{fluideq}) and the equation of motion for the order parameter (\ref{eq_sigma}) for these
perturbations, we obtain the following matrix equation,
\[
\mathcal{A}\delta X=0,
\]
where $\delta X^{T}=(\delta T(\omega ,k),\delta u^{x}(\omega ,k),\delta \sigma (\omega ,k))$ and
\[
\mathcal{A}=\left(
\begin{array}{ccc}
i\omega \left( \frac{\partial \varepsilon (T,\sigma )}{\partial T}\right) _{0} & -ik(\varepsilon _{0}+P_{0}) & i\omega \left[ \left( \frac{\partial
\varepsilon (T,\sigma )}{\partial \sigma }\right) _{0}-g\rho _{s}(\sigma _{0},T_{0})\right]  \\ -ik\left( \frac{\partial P(T,\sigma )}{\partial
T}\right) _{0} & i\omega (\varepsilon _{0}+P_{0}) & -ik\left[ g\rho _{s}(\sigma _{0},T_{0})+\left( \frac{\partial P(T,\sigma )}{\partial \sigma
}\right) _{0}\right]  \\ g\left( \frac{\partial \rho _{s}(\sigma ,T)}{\partial T}\right) _{0} & 0 & -(\omega ^{2}-k^{2})+m^{2}(\sigma _{0},T_{0})
\end{array}
\right) .
\]
The dispersion relations for perturbations are obtained by solving the equation
\begin{equation}
\mathrm{det}[\mathcal{A}]=0.
\end{equation}

As an example, when there is no coupling between the quark fluid and the chiral order parameter, that is, $g=0$, the dispersion relations are given by
\begin{eqnarray}
\omega ^{2} &=&\left( \frac{\partial P}{\partial \varepsilon }\right) _{\sigma _{0}}k^{2}~,  \label{soundm} \\ \omega ^{2} &=&k^{2}+\lambda
^{2}(3\sigma _{0}^{2}-v^{2})~.  \label{chiralm}
\end{eqnarray}
The physical interpretation of these equations is rather obvious: the first solution describes the sound wave in the quark fluid, while the second one
gives the dispersion relation for the sigma field fluctuations. As was discussed earlier, $(\partial P/\partial \varepsilon )_{\sigma _{0}}$ can be
negative at the first order phase transition. In this case, we obtain solutions with $\omega ^{2}<0$ or $\omega =\pm i|\omega |$, i.e., one of the
solutions has negative imaginary part. Such perturbations will grow exponentially, signaling the instability of the homogeneous static state.

For finite $g$, the dispersion relations are given by
\begin{equation}
\omega ^{2}=\frac{-b\pm \sqrt{b^{2}-4ac}}{2a}.  \label{disp}
\end{equation}
where
\begin{eqnarray}
a &=&\left( \frac{\partial \varepsilon }{\partial T}\right) _{0}~, \\ b &=&-k^{2}\left[ \left( \frac{\partial \varepsilon }{\partial T}\right)
_{0}+\left( \frac{\partial P}{\partial T}\right) _{0}\right] -m_{\sigma }^{2}(\sigma _{0},T_{0})\left( \frac{\partial \varepsilon }{\partial T}
\right) _{0}-\left[ \left( \frac{\partial \varepsilon }{\partial \sigma } \right) _{0}-g\rho _{s}(\sigma _{0},T_{0})\right] g\left( \frac{\partial
\rho _{s}}{\partial T}\right) _{0}~, \\ c &=&k^{4}\left( \frac{\partial P}{\partial T}\right) _{0}+k^{2}m_{\sigma }^{2}(\sigma _{0},T_{0})\left(
\frac{\partial P}{\partial T}\right) _{0}-k^{2}\left[ \left( \frac{\partial P}{\partial \sigma }\right) _{0}+g\rho _{s}(\sigma _{0},T_{0})\right]
g\left( \frac{\partial \rho _{s}}{\partial T}\right) _{0}~,
\end{eqnarray}
and $m_{\sigma }^{2}(\sigma _{0},T_{0})$ is defined in Eq.\ (\ref{sigmass}).

In general, Eq.\ (\ref{disp}) has four different solutions. However, they are symmetric with respect to the axis of $\omega =0$. Thus we show only the
positive branches of $\mathrm{Re}~\omega $ and $\mathrm{Im}~\omega $. The solution passing through the origin, $(\omega ,k)=(0,0)$ (analogous to Eq.\
(\ref{soundm})), is associated with the propagation of sound waves and is called the sound branch. The other solution, which has a mass gap at $k=0$,
corresponds to the propagation of the sigma field fluctuations and is called the sigma branch. Using Eqs.\ (\ref{disp}) one can obtain the following
explicit expression for the mass gap:
\begin{equation}
m_{\mathrm{pole}}^{2}(\sigma _{0},T)\equiv \omega ^{2}(k\rightarrow 0)=m_{\sigma }^{2}(\sigma _{0},T_{0})-g\left( \frac{\partial \rho _{s}}{\partial T}\right) _{0}\left( \frac{\partial s}{\partial \sigma }\right)
_{0}\left( \frac{\partial s}{\partial T}\right) _{0}^{-1}~.  \label{massgap}
\end{equation}
Here the second term comes from the sigma coupling to the sound branch. It is easy to check that this term is positive. Thus, the sigma excitations in
the chiral fluid are characterized by the "pole mass" (\ref{massgap}), which differs from the "screening mass" (\ref{sigmass}). In high temperature
limit one can obtain the following relation
\begin{equation}
m_{\mathrm{gap}}^{2}-m_{\sigma }^{2}=\nu _{q}g^{2}\frac{3M^{2}}{4\pi ^{2}}\left[ \ln {\left( \frac{M}{\pi T}\right) }+\gamma -\frac{1}{2}\right] ~,
\end{equation}
where $M=g\sigma _{0}$ is the constituent quark mass and $\gamma \approx 0.577$ is Euler's constant. In the vacuum, these two masses coincide, $m_{\sigma }=m_{\mathrm{gap}}\approx \sqrt{2}\lambda f_{\pi }$, see Eq.\ (\ref{chiralm}).

\begin{figure}[h]
\includegraphics[scale=0.3]{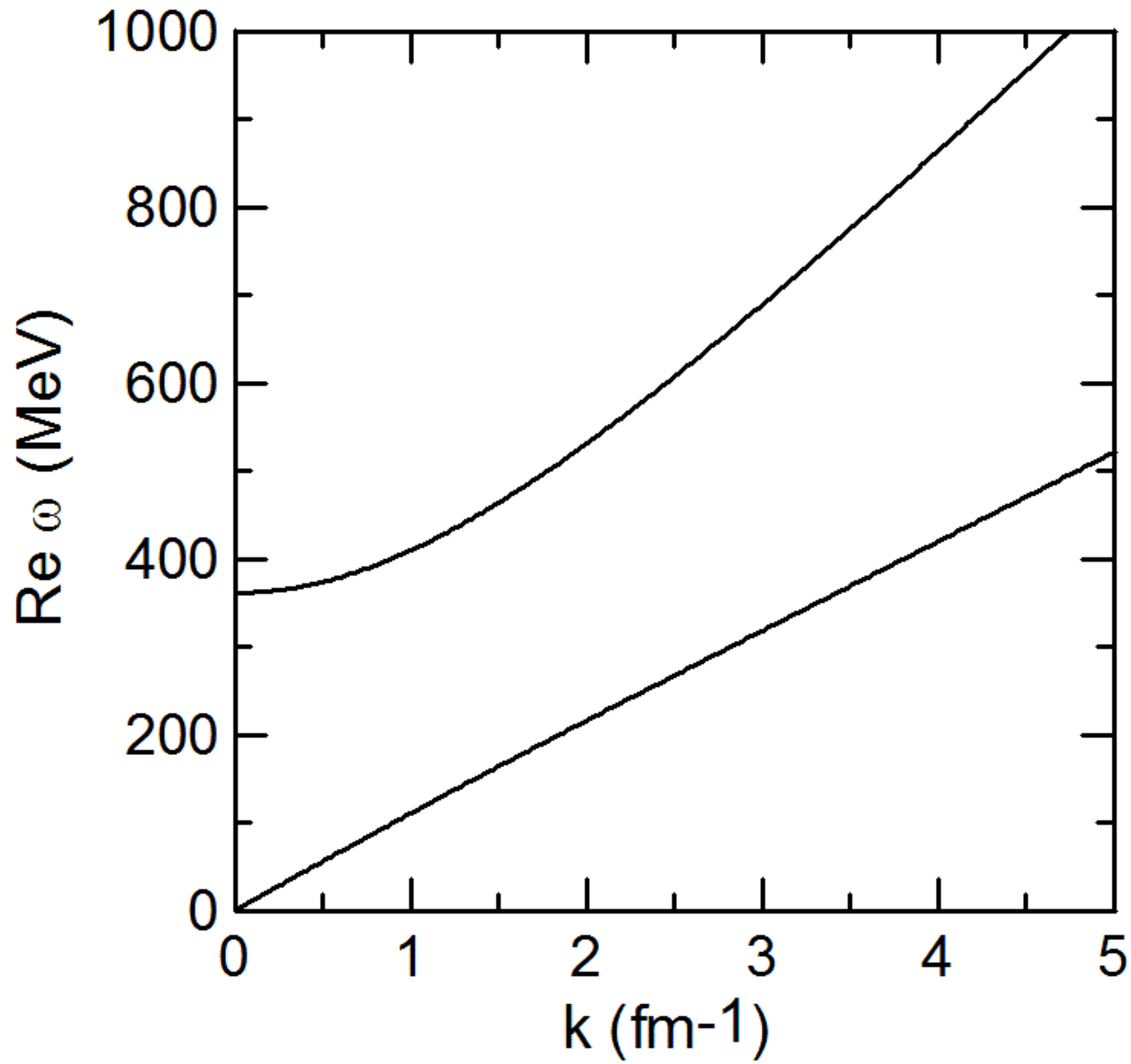} %
\includegraphics[scale=0.3]{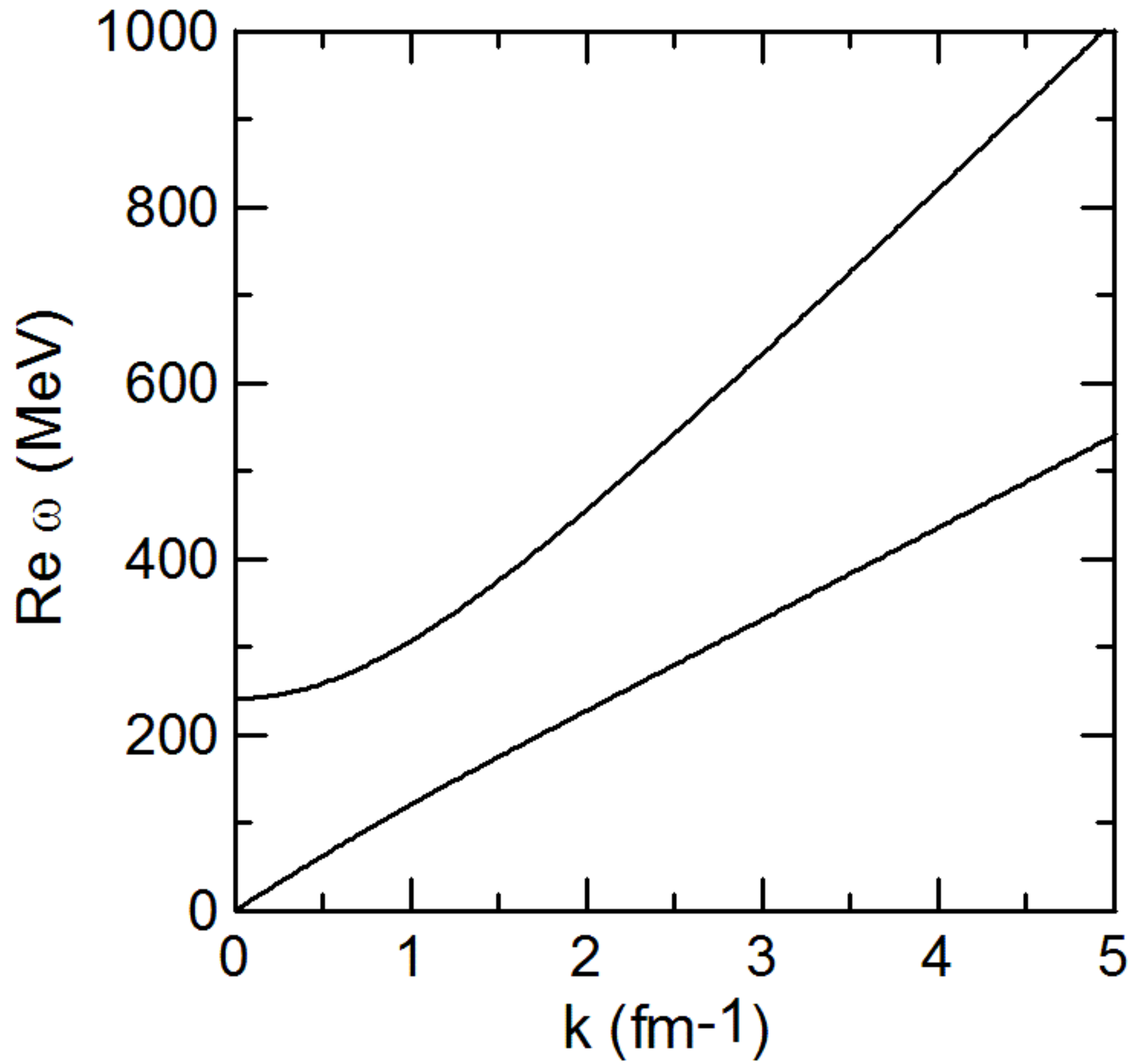} %
\includegraphics[scale=0.3]{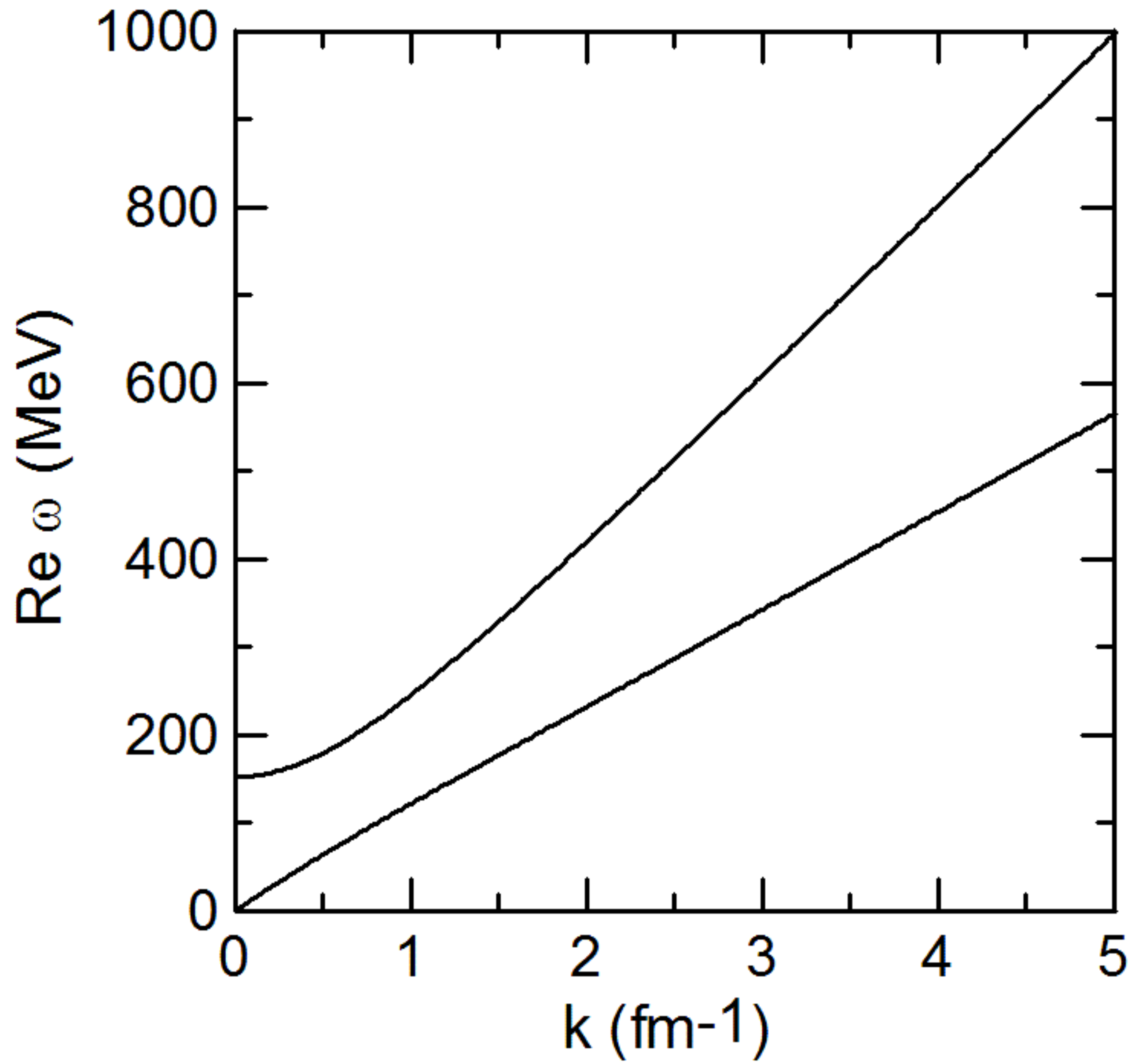} %
\includegraphics[scale=0.3]{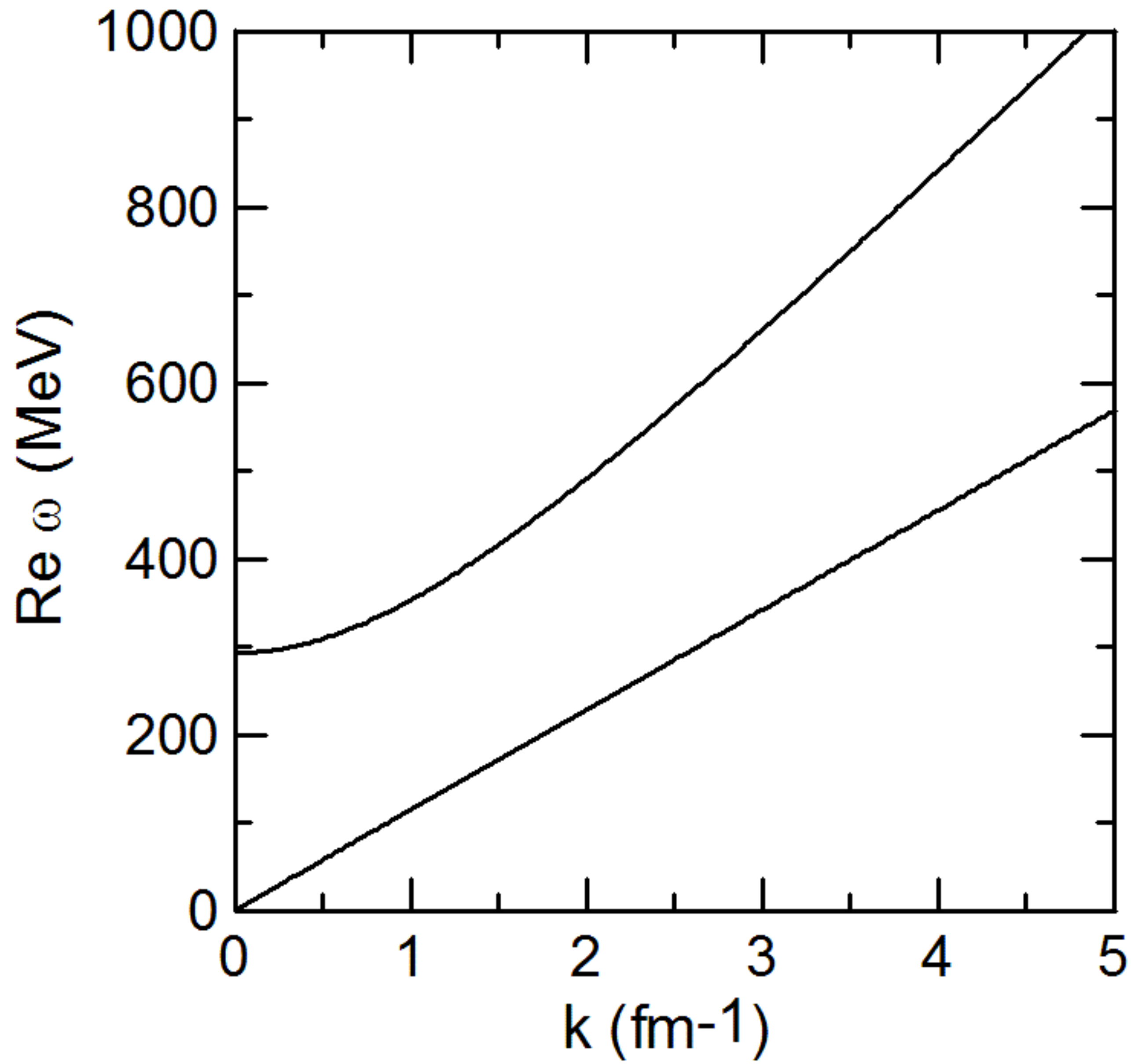}
\caption{The real part of the dispersion relation for $g=3.3$ at $T=130$ (left up), $140$ (right up) and $150$ (left down) MeV and $160$ MeV (right
down), respectively. For $g=3.3$, there is no imaginary part. } \label{dispers3.3}
\end{figure}

In case of the crossover transition, $g=3.3$, there is no imaginary part and we show only the real parts of the dispersion relations for four different
temperatures, $130$ (left up), $140$ (right up), $150$ (left down) and $160$ (right down) MeV. These temperatures
which cover the region from the
chirally broken phase to the restored phase. In this case the two branches are always separated from each other and never intersect. It is interesting
to note that the mass gap for the sigma branch first goes down with $T$, reaches a minimum at $T\approx 150$ MeV and then grows again. This is fully
consistent with the results obtained in Ref.\ \cite{Scavenius1} for $g=3.3$ and small $\mu $.

\begin{figure}[tbp]
\includegraphics[scale=0.3]{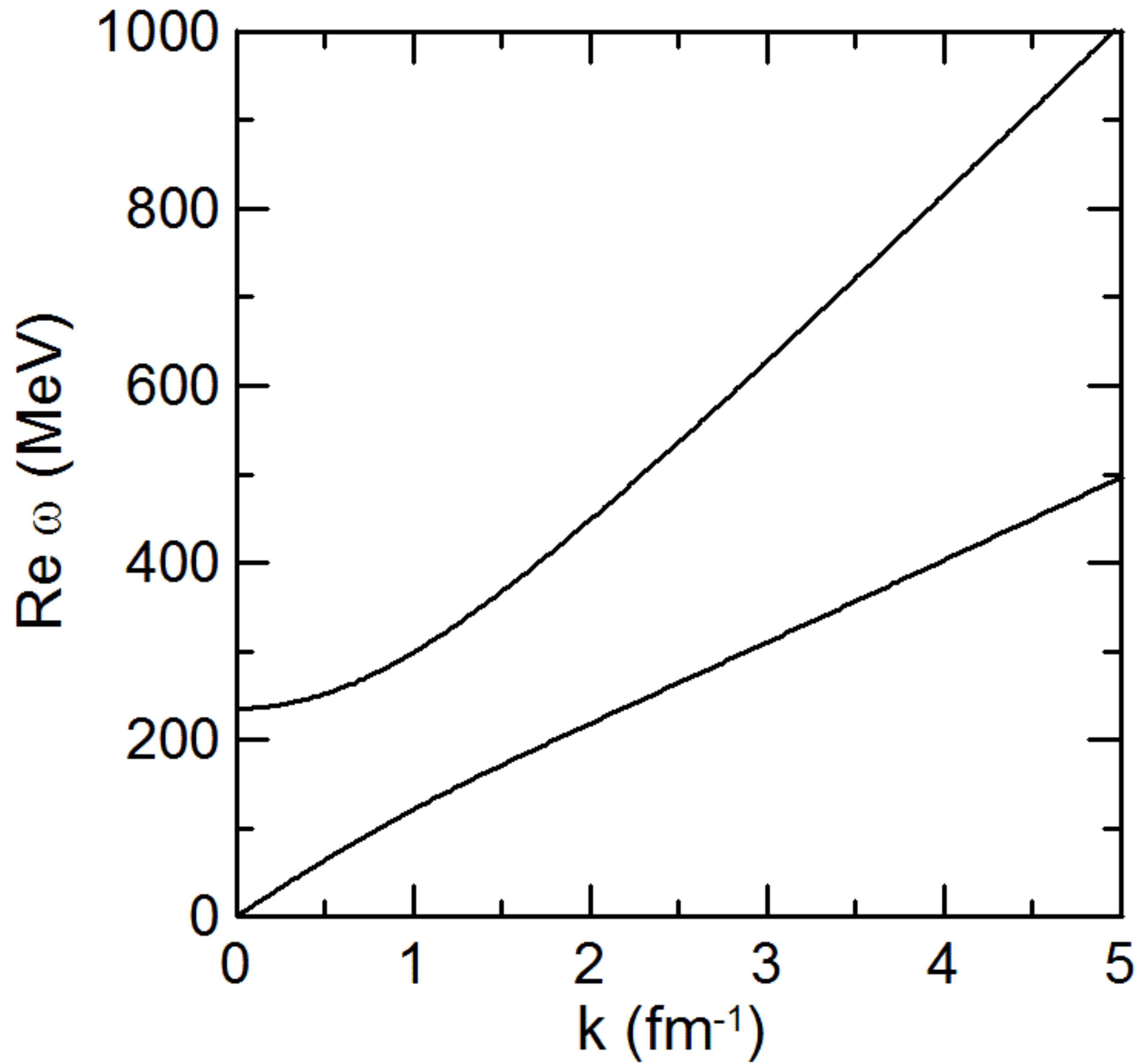} %
\includegraphics[scale=0.3]{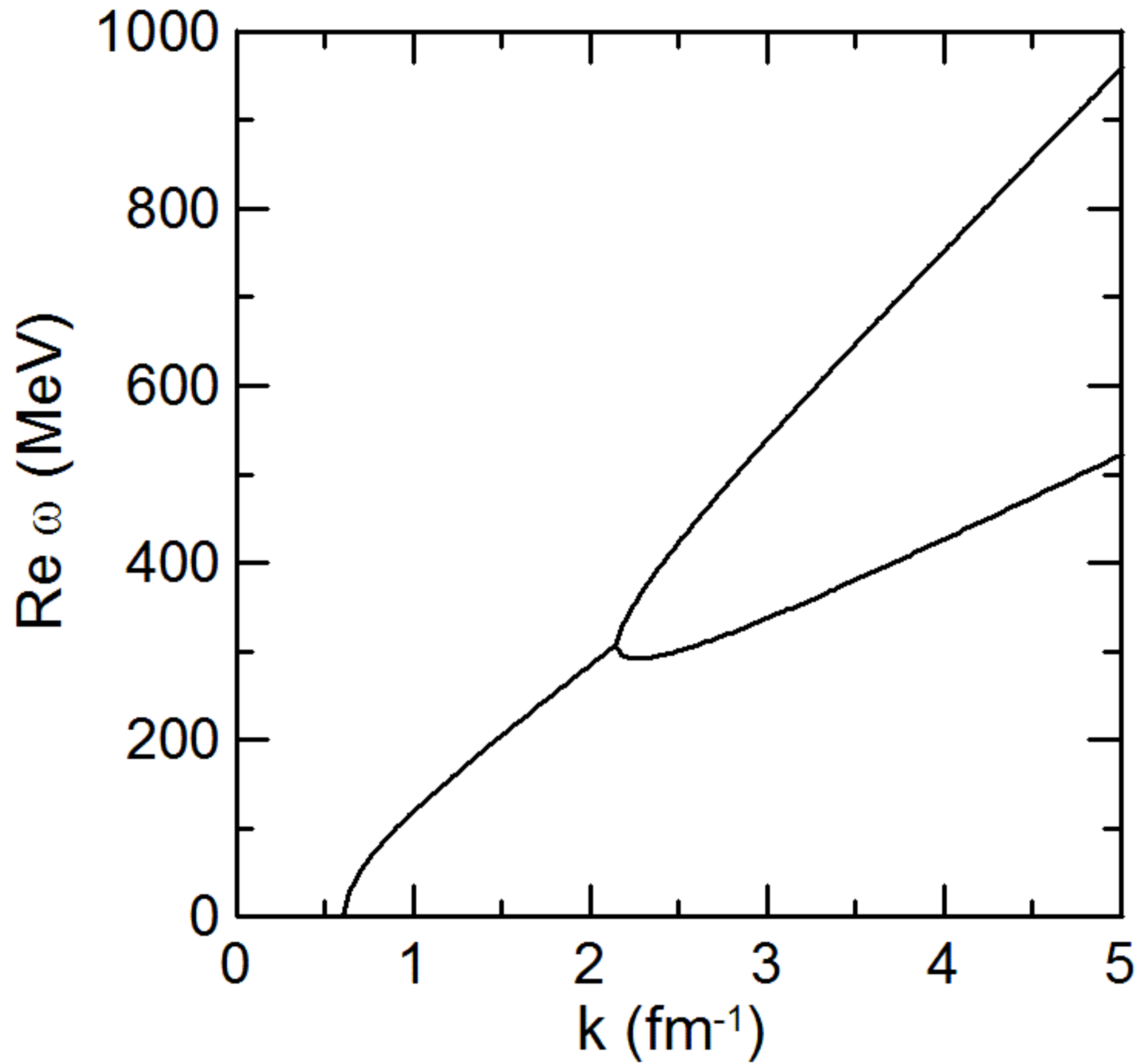} %
\includegraphics[scale=0.3]{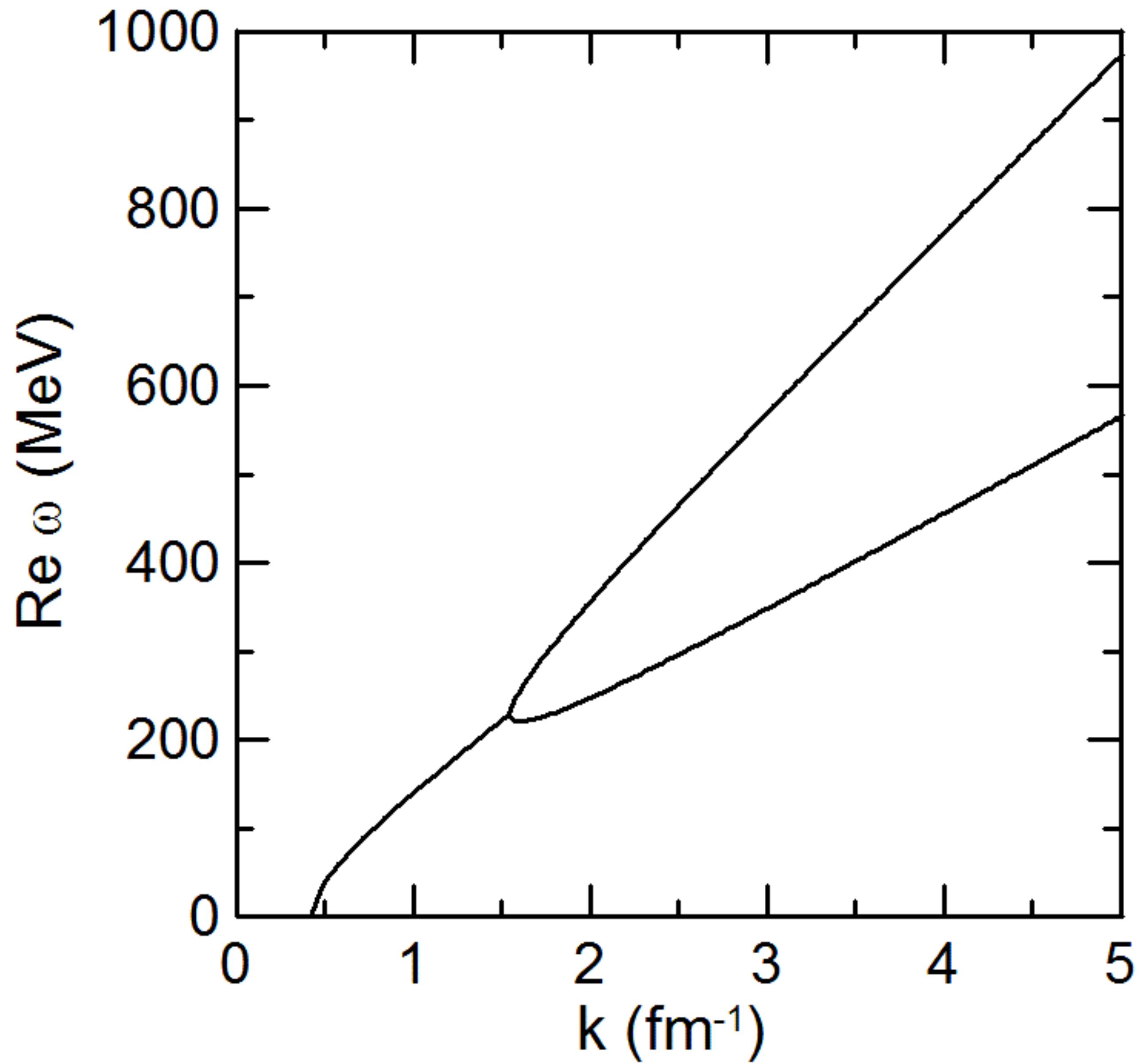} %
\includegraphics[scale=0.3]{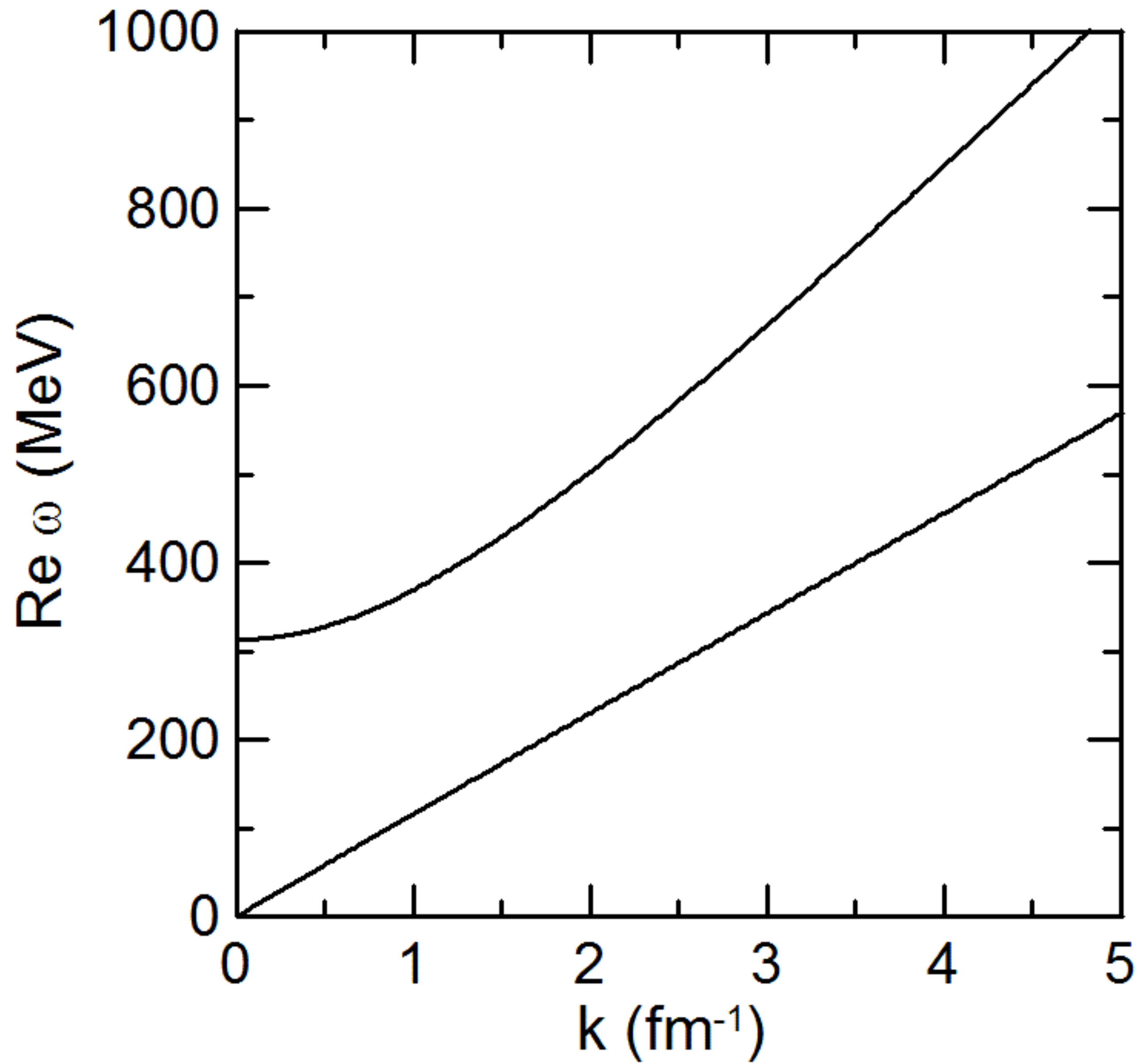}
\caption{Real parts of the dispersion relation for $g=4.5$ at the point A (left up), B (right up), D (left down) and E (right down), respectively.}
\label{dispersre4.5}
\end{figure}

\begin{figure}[tbp]
\includegraphics[scale=0.3]{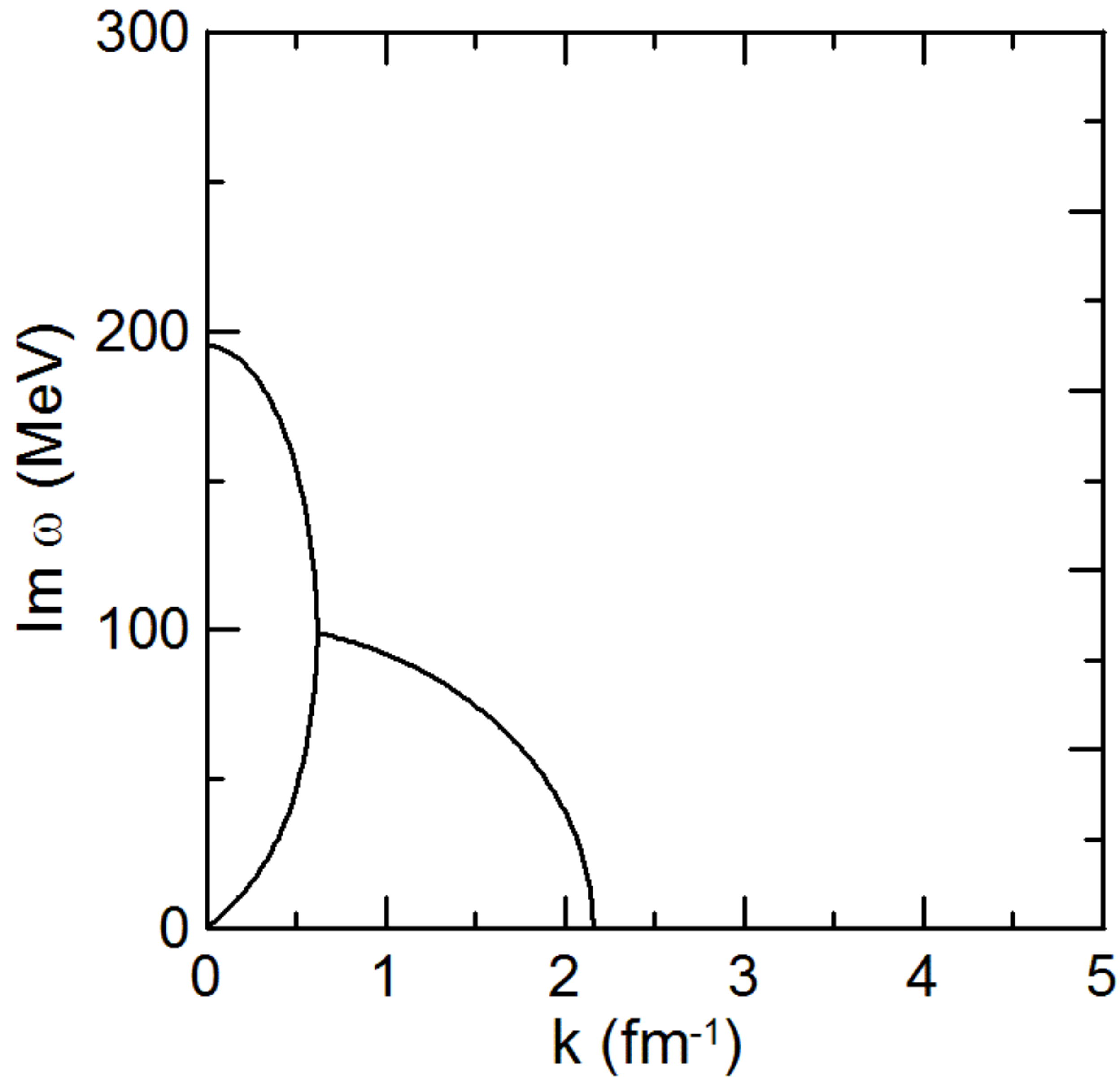} %
\includegraphics[scale=0.3]{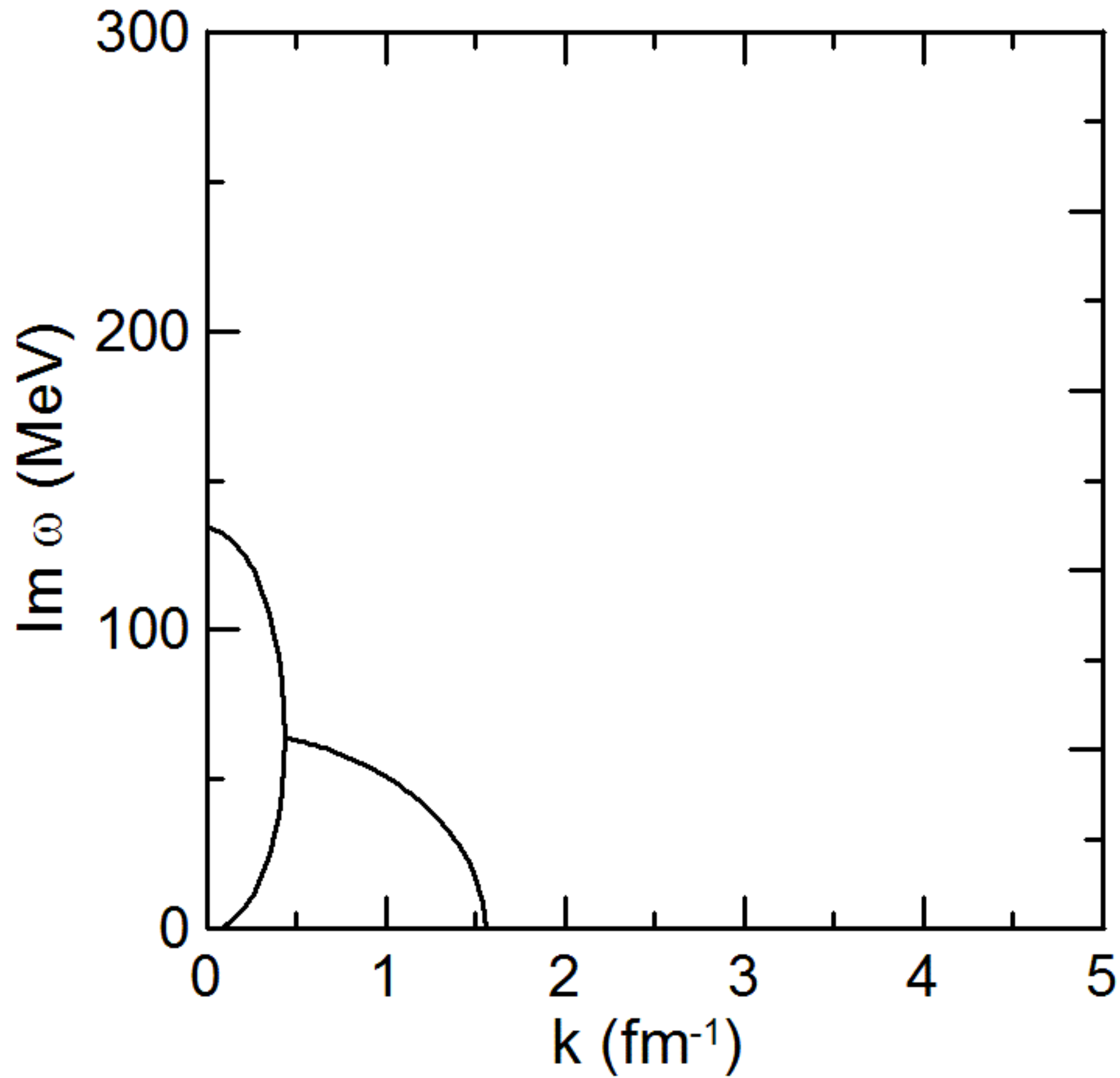}
\caption{Imaginary parts of the dispersion relation for $g=4.5$ at points B (left) and D (right), respectively. The imaginary parts at the points A and
E vanish.} \label{dispersim4.5}
\end{figure}

The behavior becomes more complex for $g=4.5$. Figures \ref{dispersre4.5} and \ref{dispersim4.5}, respectively, show the real and imaginary parts of
the dispersion relation for different situations corresponding to the four points A, B, D and E in Fig.\ \ref{gapfig}. In these calculations we did not
apply the Maxwell construction
because we consider the quenching of the background of the hydrodynamic variables.
but also considered metastable
states on segments AB and DE. As has been mentioned earlier, in the temperature interval between $T_{\mathrm{D}}=122.7$ MeV and $T_{\mathrm{B}}=132.1$
MeV, there exists three solutions for the sigma field, corresponding to two local minima of the thermodynamic potential, and one maximum in between.
For the background states which have temperatures $T>T_{\mathrm{B}}$ or $T<T_{\mathrm{D}}$, the thermodynamic potential has only one local minimum. In
this case the two branches of the dispersion relation are separated and have no imaginary parts, as shown for the points A and E in Fig.\
\ref{dispersre4.5}.


The states along the line BCD correspond to a local maximum of the thermodynamic potential and the perturbations around these states are trivially
unstable (spinodal instability). On the other hand, a non-trivial behavior is observed for the metastable states on segments AB ($T_{\mathrm{C}}<T<T_{\mathrm{B}}$) and DE ($T_{\mathrm{D}}<T<T_{\mathrm{C}}$). Although these states represent local minima of the thermodynamic potential, $c_{s}^{2}$ becomes negative in this region of temperatures which induces an instability through the coupling to the hydrodynamic modes. In order to
illustrate this point, the real and imaginary parts of the dispersion relation for the points D and B are shown in Figs.\ \ref{dispersre4.5} and
\ref{dispersim4.5}. One can see that at $k<k_{\mathrm{max}}\sim 2$ fm$^{-1}$ the sound and sigma branches degenerate into a single branch (Fig.\ \ref
{dispersre4.5}) which has a positive imaginary part (Fig.\ \ref{dispersim4.5}). It is interesting to note that for $g=4.5$, both branches of the
dispersion relation become unstable at $k<k_{\mathrm{max}}$, i.e. when either $c_{s}^{2}<0$ (sound waves) or $m_{\mathrm{pole}}^{2}<0$ (sigma waves).

The appearance of the growing modes of density fluctuations, called as spinodal instability, is very well known in the physics of first order phase transitions, see e.g reviews \cite{Binder,Pethick,Chomaz}. This instability develops when a homogeneous system is suddenly quenched into the spinodal region of the phase diagram, defined by the condition $c_s^2<0$. Usually the instability region is limited by moderate wave numbers, $k<k_{\mathrm{max}}$, as is clearly seen in Fig.\ \ref{dispersim4.5}.
The value of $k_{\mathrm{max}}$ is determined by the non-locality scale of the particle interaction, which is given by $|m_{\mathrm{pole}}|$ in our
model.

\section{Evolution of fluctuations in a dynamical background \label{secbjorken}}

Usually the dynamics of a first order phase transition is discussed in the framework of the nucleation theory, assuming the thermal activation or
penetration through the barrier separating the two phases. However, one should bear in mind that, in a rapidly expanding system, the thermodynamic
potential is changing with the characteristic expansion time $\tau_{\mathrm{exp}}$. If $\tau_{\mathrm{exp}}$ is much shorter than the barrier
penetration time, the nucleation process will not be efficient to fulfill the transition. Then the formation of the new phase will start only when the
barrier disappears and the system may freely roll down toward a more thermodynamically stable state \cite{mish1,mish2}. That is, the phase transition
mechanism changes form the nucleation to the spinodal decomposition. In this section we discuss the dynamics of fluctuations in a rapidly expanding
background of the Bjorken type \cite{bjorken}.

\subsection{Boost-invariant background solution}

The hot matter created in relativistic heavy-ion collisions exhibits a strong collective expansion, predominantly in the longitudinal direction of the
collision. For the sake of simplicity, we consider a (1+1)-dimensional system in $\tau$-$\eta$ coordinates defined by
\begin{eqnarray}
\tau = \sqrt{t^2 - z^2},~~~~~\eta = \frac{1}{2}\ln\left[ \frac{t+z}{t-z}\right],
\end{eqnarray}
and parameterize the fluid velocity as
\begin{eqnarray}
u^{\mu} = (\cosh \theta , \sinh \theta).
\end{eqnarray}

By using these variables, we obtain the following relations,
\begin{eqnarray}
\left(
\begin{array}{c}
\partial_0 \\
\partial_1
\end{array}
\right) = \left(
\begin{array}{cc}
\cosh \eta & - \sinh \eta \\ -\sinh \eta & \cosh \eta
\end{array}
\right) \left(
\begin{array}{c}
\partial_{\tau} \\
\frac{1}{\tau}\partial_{\eta}
\end{array}
\right)~.
\end{eqnarray}
It is convenient to introduce the following notations,
\begin{eqnarray}
\left(
\begin{array}{c}
D \\ \nabla
\end{array}
\right) = \left(
\begin{array}{cc}
\cosh (\theta - \eta) & \sinh (\theta - \eta) \\ \sinh (\theta -\eta) & \cosh (\theta - \eta)
\end{array}
\right) \left(
\begin{array}{c}
\partial_{\tau} \\
\frac{1}{\tau}\partial_{\eta}
\end{array}
\right)~.
\end{eqnarray}
Then one can write
\begin{equation}
u^{\mu}\partial_{\mu} = D,~~~~\partial_{\mu}u^{\mu} = \nabla \theta.
\end{equation}
Now the hydrodynamic equations (\ref{fluideq}) can be written as
\begin{eqnarray}
&& D\varepsilon + (\varepsilon + P)\nabla \theta - g\rho_s D \sigma = 0~, \label{hydroeq-1} \\ && (\varepsilon + P)D \theta + \nabla P + g\rho_s \nabla
\sigma = 0~. \label{hydroeq-2}
\end{eqnarray}
Using the generalized thermodynamics relations, Eqs.\ (\ref{eq_GD}) and (\ref{1st_law}), we can reduce these equations to a more compact form,
\begin{eqnarray}
&& D s + s \nabla \theta = 0~,  \label{hydroeq-1a} \\ && D \theta + \nabla \ln T = 0~.  \label{hydroeq-2a}
\end{eqnarray}
It is easy to show that these equations are consistent with the entropy conservation equation, $\partial_\mu(su^\mu)=0$.

For the background solution, we take the boost-invariant ansatz proposed by Bjorken \cite{bjorken}, i.e. we set $\eta = \theta$ and assume that all
hydrodynamic variables are functions of $\tau$ only. Then the hydrodynamic equations (\ref{hydroeq-1a}) and (\ref{hydroeq-2a}) are reduced to
\begin{eqnarray}
&& T ( s_T \partial_\tau T + s_{\sigma} \partial_\tau \sigma ) + \frac{Ts}{\tau} = 0,  \label{hydro1}
\end{eqnarray}
where $s_T = \left( \partial s/\partial T \right)_\sigma , s_\sigma = \left(
\partial s/\partial \sigma \right)_T $. On the other hand, the equation of
motion for the chiral order parameter is
\begin{equation}  \label{eom-2}
\partial^2_\tau \sigma + \frac{1}{\tau} \partial_\tau \sigma + \lambda^2 (
\sigma^2 - v^2 )\sigma - H =-g \rho_s~.
\end{equation}

\begin{figure}[tbp]
\includegraphics[scale=0.3]{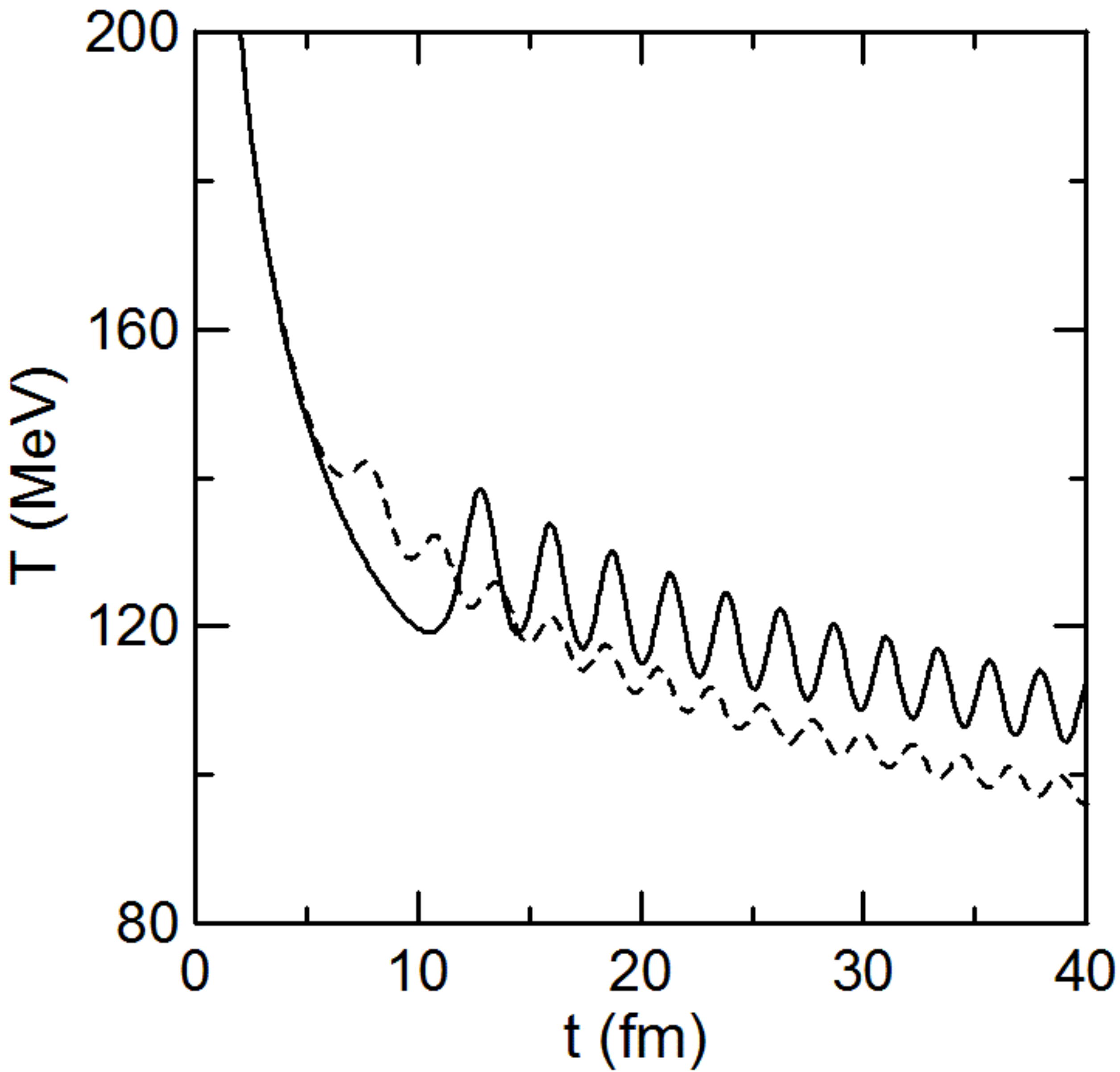} %
\includegraphics[scale=0.3]{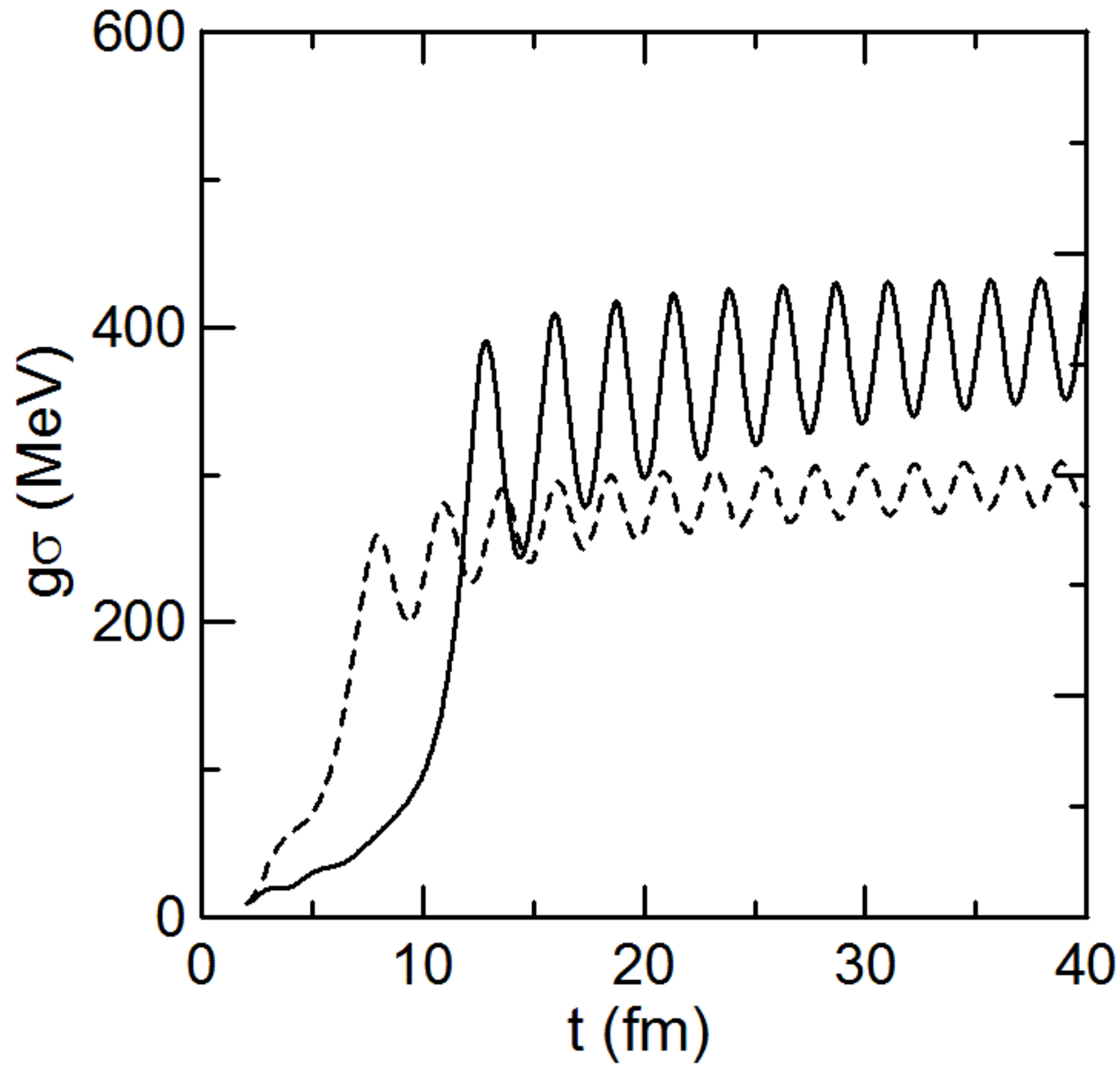}
\caption{The temperature $T$ (left panel) and the chiral order parameter $g\protect\sigma$ (right panel) as functions of proper time in a
boost-invariant solution. The dashed and solid lines represent the results for $g=3.3$ and $4.5$, respectively. } \label{aveboost}
\end{figure}

The time dependencies of the temperature $T$ and the chiral order parameter $g\sigma$ are shown in Fig.\ \ref{aveboost} for the initial conditions
set at the initial time $\tau_0 = 2$ fm: $T=200$ MeV $g \sigma = 10$ MeV and
$\partial_\tau \sigma = 0$. The dashed and solid lines represent the results
for $g=3.3$ and $4.5$, respectively. There is a qualitative difference in the behavior of the two lines, namely, the time delay in the first order
phase transition (solid line) as compared with the crossover (dashed line). This difference is more evident in the time evolution of $g\sigma$. The
time delay can be explained by the trapping of the order parameter field in the metastable minimum during the initial stage of expansion, until it
disappears at point D (see Fig.\ \ref{gapfig}). As a matter of fact, the values of $g\sigma$ at $\tau \approx 10$ fm correspond to metastable states
around point D. This is a supercooling process which occurs at a fast expansion through the first order phase transition. After this point is reached,
the potential barrier disappears and the field rolls down toward the stable minimum. Then it oscillates there for a long time, because of the lack of
true dissipation except of the expansion.

\subsection{Stability analysis in boost-invariant background}

In order to investigate stability of the boost-invariant solution, we consider plane-wave perturbations of $\delta \sigma$ and other quantities, but
now only in the $\eta$ coordinate, that is, we study the perturbations of the type
\begin{equation}
\delta\sigma(\eta,\tau)=\delta\sigma_k(\tau) e^{-ik\eta}~.
\end{equation}
Then expanding equations (\ref{hydroeq-1a}), (\ref{hydroeq-2a}) and (\ref{eom-2}) to the linear order in perturbations we obtain the following
time-dependent matrix equation,
\begin{equation}
\partial_\tau \delta X_k(\tau) = \mathcal{A}_k(\tau) \delta X_k(\tau),
\label{scaling_sta}
\end{equation}
where $\delta X^T_k(\tau) = \left\{\delta T_k (\tau), \partial_\tau \delta \sigma_k(\tau), \delta\sigma_k(\tau), \delta \theta_k(\tau)\right\} $ and
\begin{equation}
\mathcal{A}_k(\tau) =\left(
\begin{array}{llll}
a_{11} & a_{12} & a_{13} & a_{14} \\ a_{21} & a_{22} & a_{23} & 0 \\ 0 & 1 & 0 & 0 \\ a_{41} & 0 & 0 & a_{44}
\end{array}
\right)~.
\end{equation}
Explicit expressions for the matrix elements are
\begin{eqnarray}
a_{11} &=& - \frac{\partial_\tau T}{T} - \frac{s_{TT} \partial_\tau T }{s_T} - \frac{s_\sigma}{T s_T}\partial_\tau \sigma -\frac{s_{\sigma T}}{s_T}
\partial_\tau \sigma - \frac{s}{\tau T s_T} -\frac{1}{\tau} , \\
a_{12} &=& -\frac{ s_\sigma }{s_T},
\\ a_{13} &=& -\frac{s_{T\sigma}}{s_T}\partial_\tau T - \frac{s_{\sigma\sigma}}{s_T} \partial_\tau \sigma -
\frac{s_\sigma}{\tau s_T}, \\
a_{14} &=& ik \frac{s}{\tau s_T}, \\ a_{21} &=& -g \rho_T ,
\\ a_{22} &=& -\frac{1}{\tau } , \\
a_{23} &=&-\frac{k^2}{\tau ^2} -\lambda^2 (3\sigma^2 - v^2) -g \rho_\sigma ,\\
a_{41} &=& \frac{ik}{\tau T} , \\ a_{44} &=& -\frac{1}{\tau } -\frac{ \partial_\tau T}{T}~,
\end{eqnarray}
where all $\tau$-dependent functions correspond to the background solution. Here we have used the short-hand notations
\begin{equation}  \label{notations}
s_T = \frac{\partial s}{\partial T},~~ s_\sigma = \frac{\partial s}{\partial \sigma},~~ s_{T\sigma}= \frac{\partial^2 s}{\partial T \partial \sigma},~~
s_{TT}= \frac{\partial^2 s}{\partial T^2},~~ s_{\sigma \sigma}= \frac{\partial^2 s}{\partial \sigma^2},~~ \rho_T = \frac{\partial \rho_s}{\partial
T},~~ \rho_\sigma = \frac{\partial \rho_s}{\partial \sigma}.
\end{equation}

Since the matrix $\mathcal{A}_{k}$ has explicit $\tau $ dependence, we cannot use the same analysis applied to the static background. In order to
investigate the stability for general parameter sets and initial conditions one can use, e.g. the Lyapunov direct method. However, the implementation
of this method for systems containing more than three variables is rather cumbersome. And even if this scheme were implemented, the stability in most
part of the parameter space would remain undetermined, see Ref.\ \cite{dkkm3}. For this reason, in this work we do not use the Lyapunov direct method
and discuss the stability only for concretely chosen parameters and initial conditions. Then we can check the stability by directly solving Eq.\ (\ref{scaling_sta}). We use the same initial values as before with the following additional conditions: $g\delta \sigma =10$ MeV, $\partial _{\tau }\delta
\sigma =0$, $\delta T=0$, $\delta \theta =0.1$.

\begin{figure}[tbp]
\hspace{-0.5cm} \includegraphics[scale=0.3]{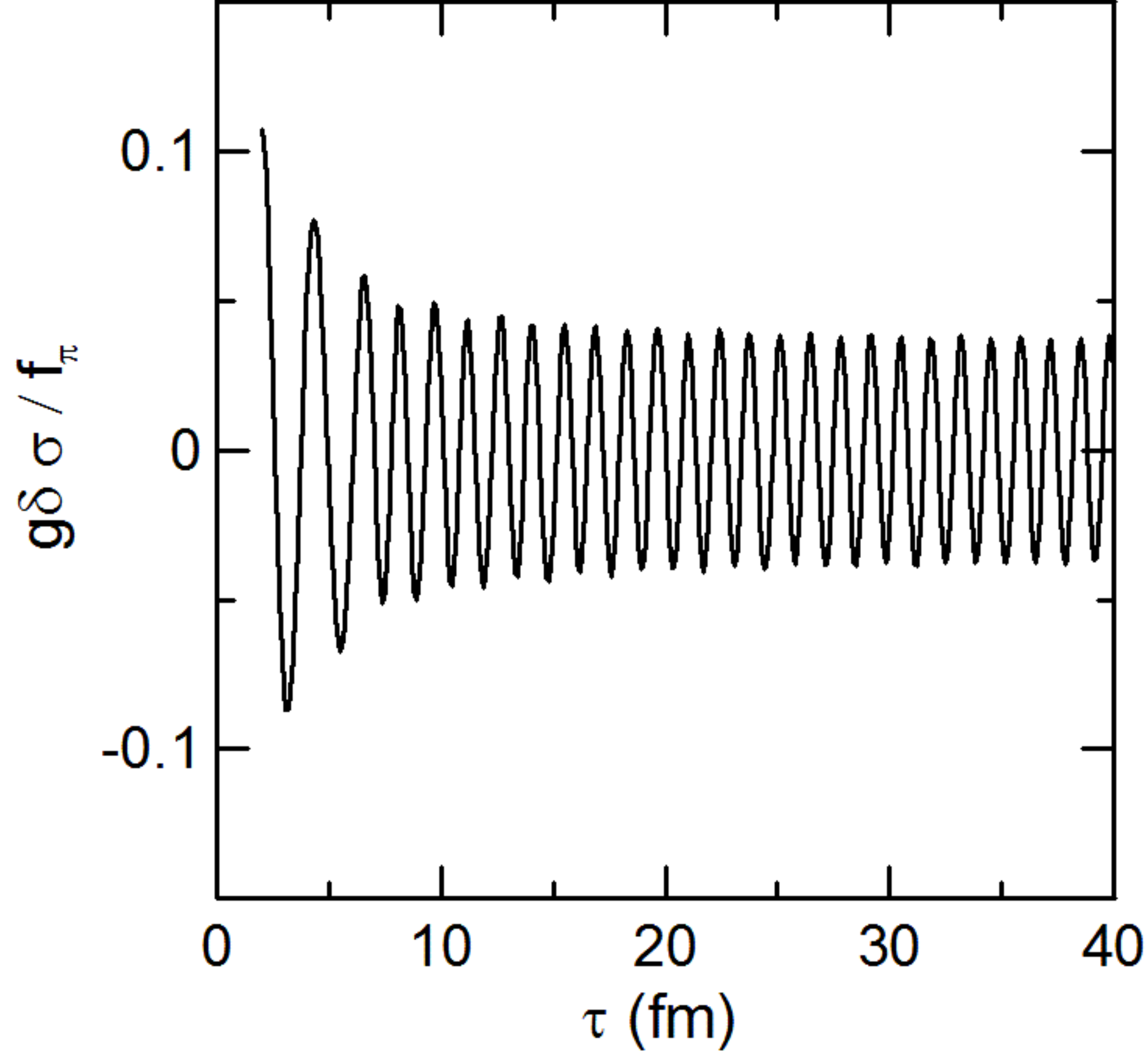}
\includegraphics[scale=0.3]{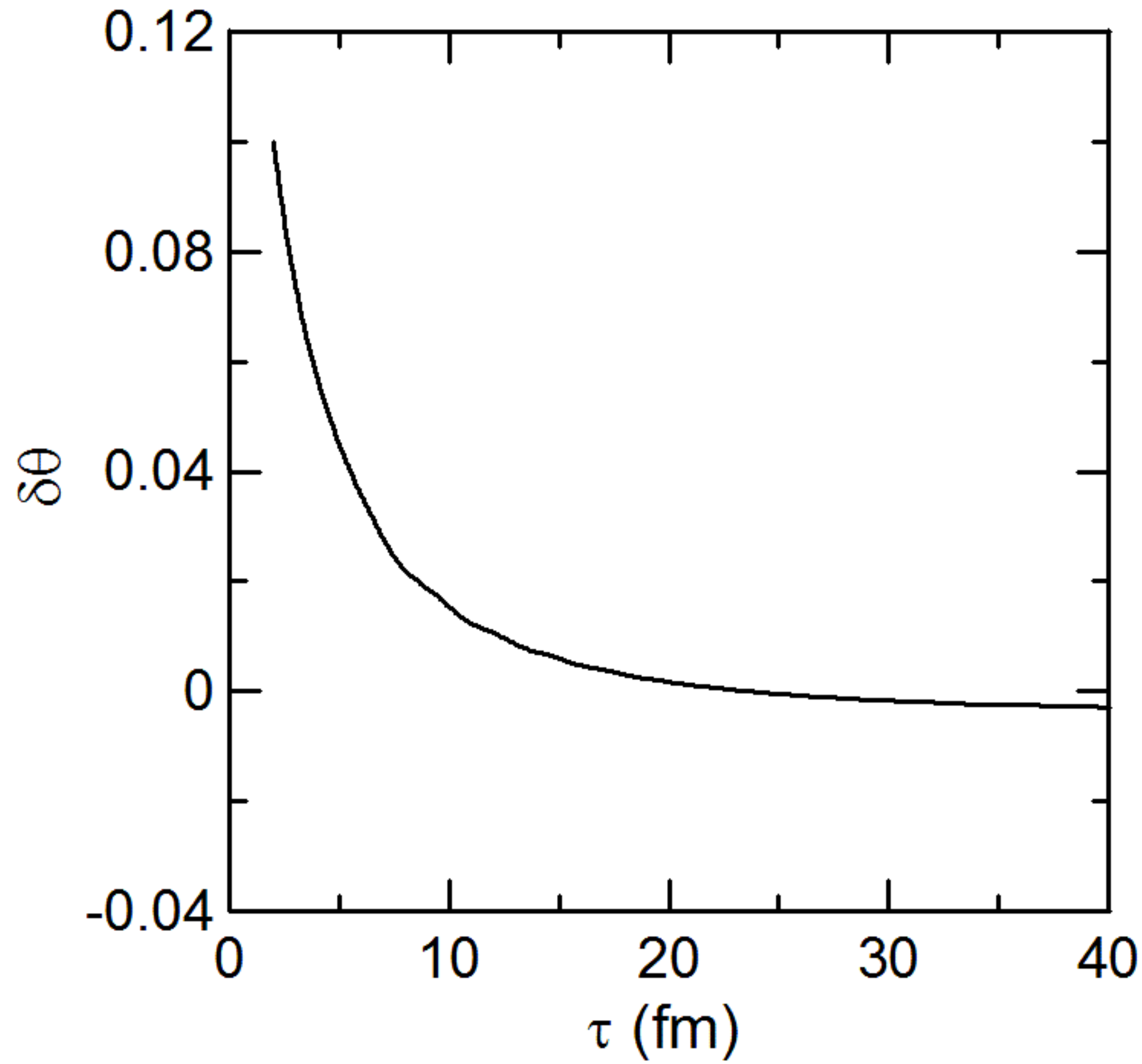} %
\includegraphics[scale=0.3]{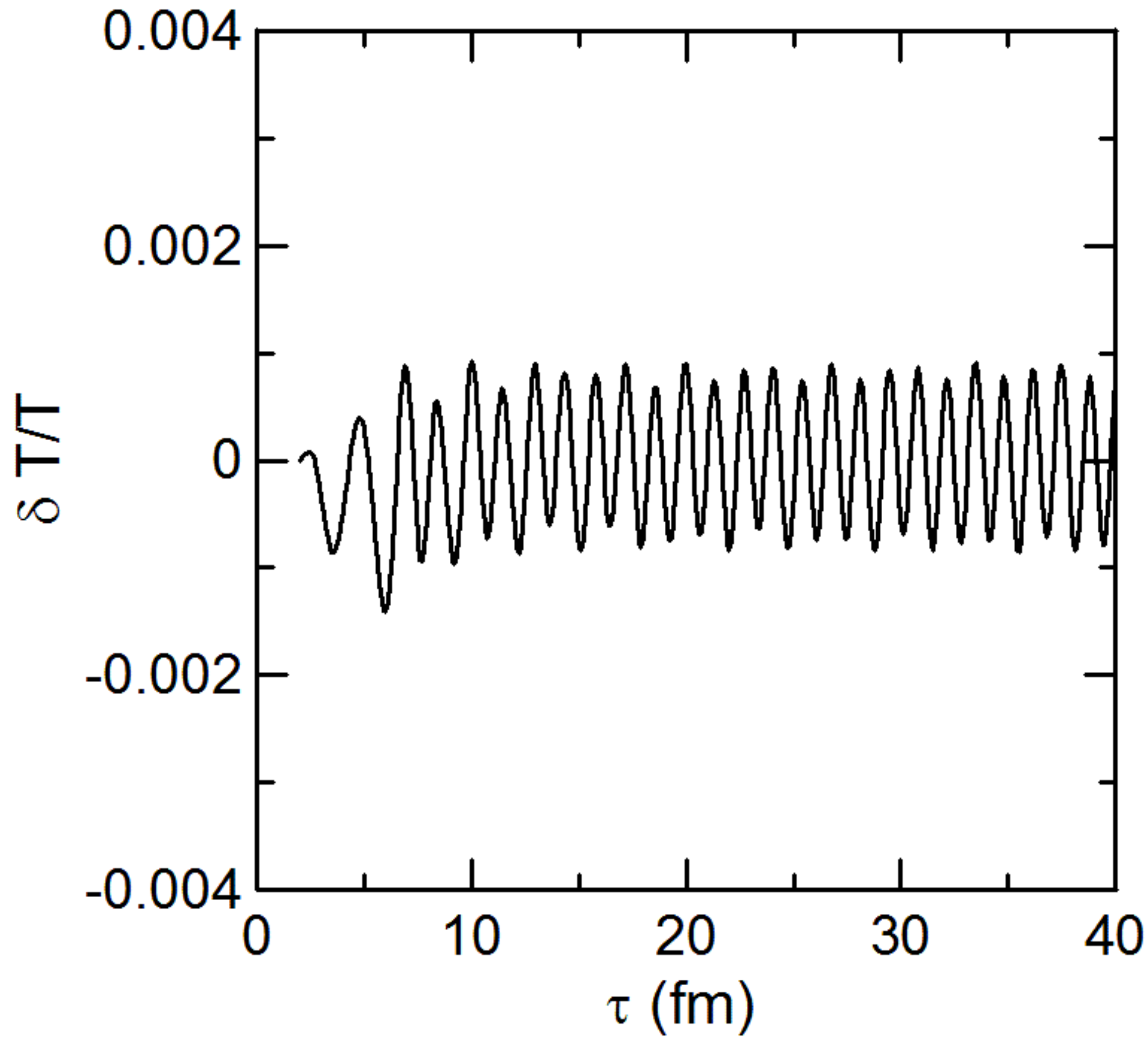}
\caption{The time dependencies of $\protect\delta \protect\sigma/\protect\sigma$ (left), $\protect\delta \protect\theta$ (middle) and $\protect\delta T/T$ (right) for $g=3.3$ at $k=1$. } \label{deltag=33}
\end{figure}

In Fig.\ \ref{deltag=33}, we show the time dependence of $g \delta \sigma/f_\pi$, $\delta T/T$ and $\delta \theta$ for $g=3.3$ and $k=1$. The
fluctuation of $\theta$ is not normalized because $\theta = 0$ at the central rapidity $\eta = 0$. One can see that the fluctuations do not increase
and, therefore, the boost-invariant solution is stable in this case. 

\begin{figure}[tbp]
\includegraphics[scale=0.3]{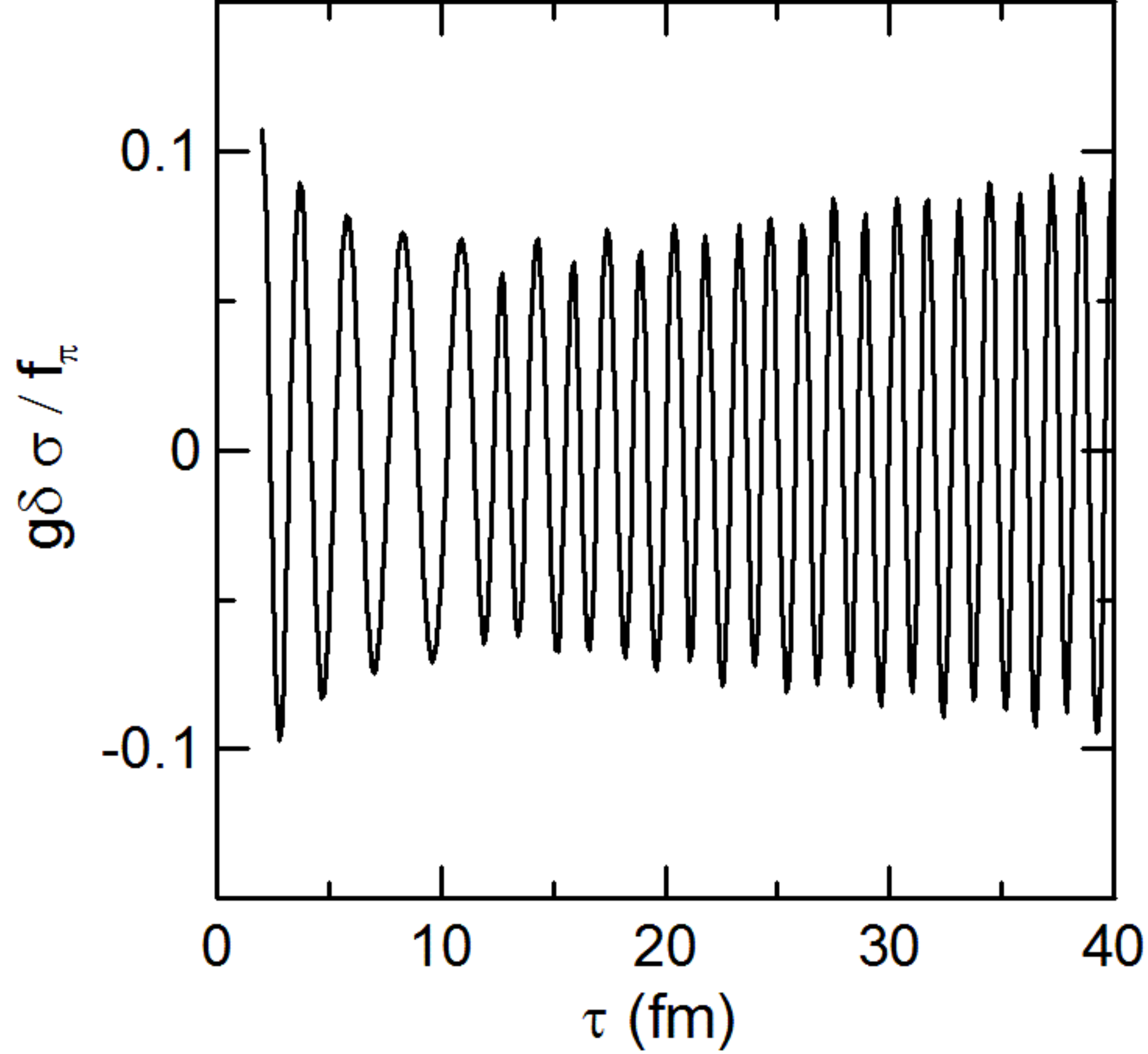}
\includegraphics[scale=0.3]{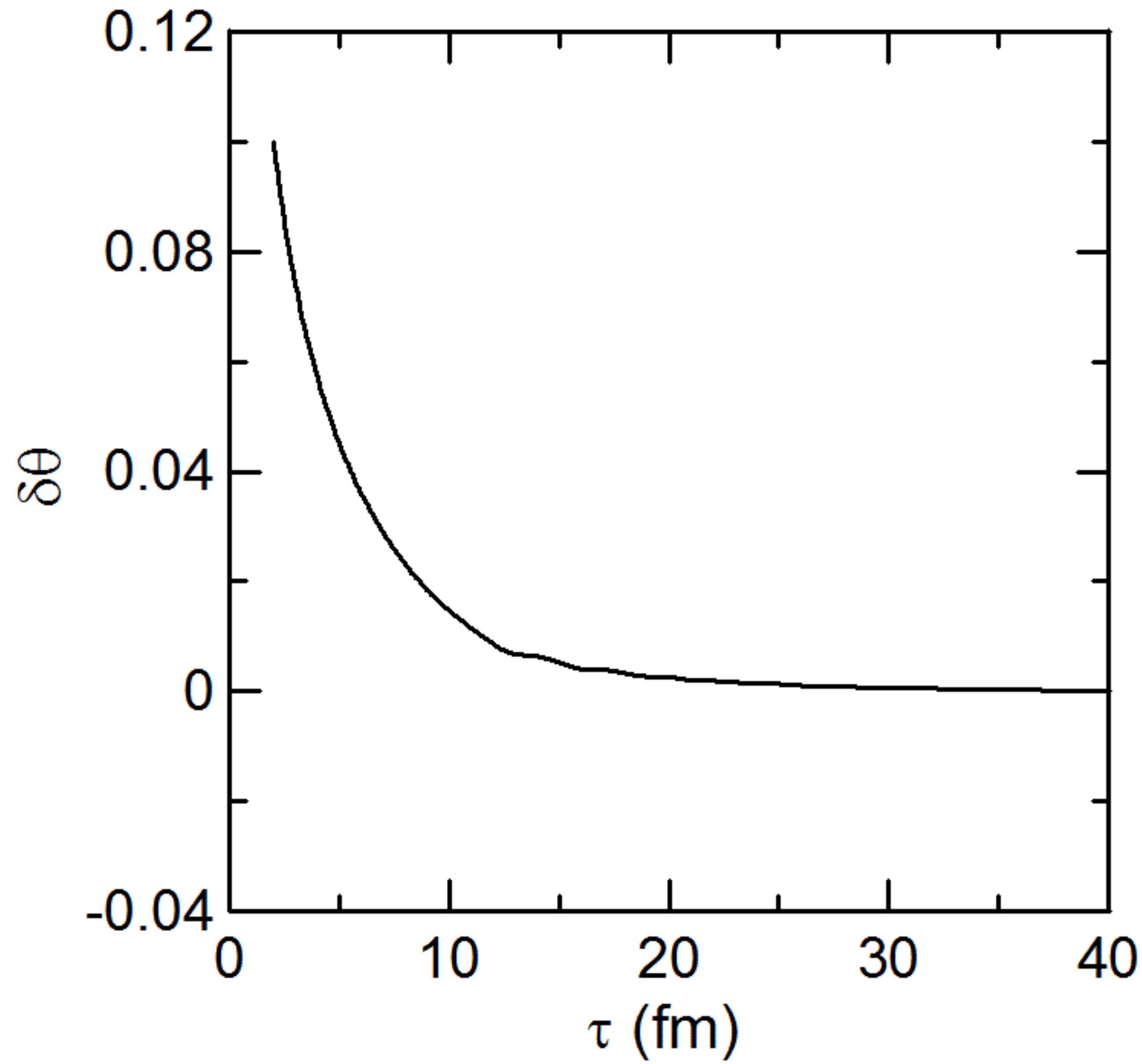}
\includegraphics[scale=0.3]{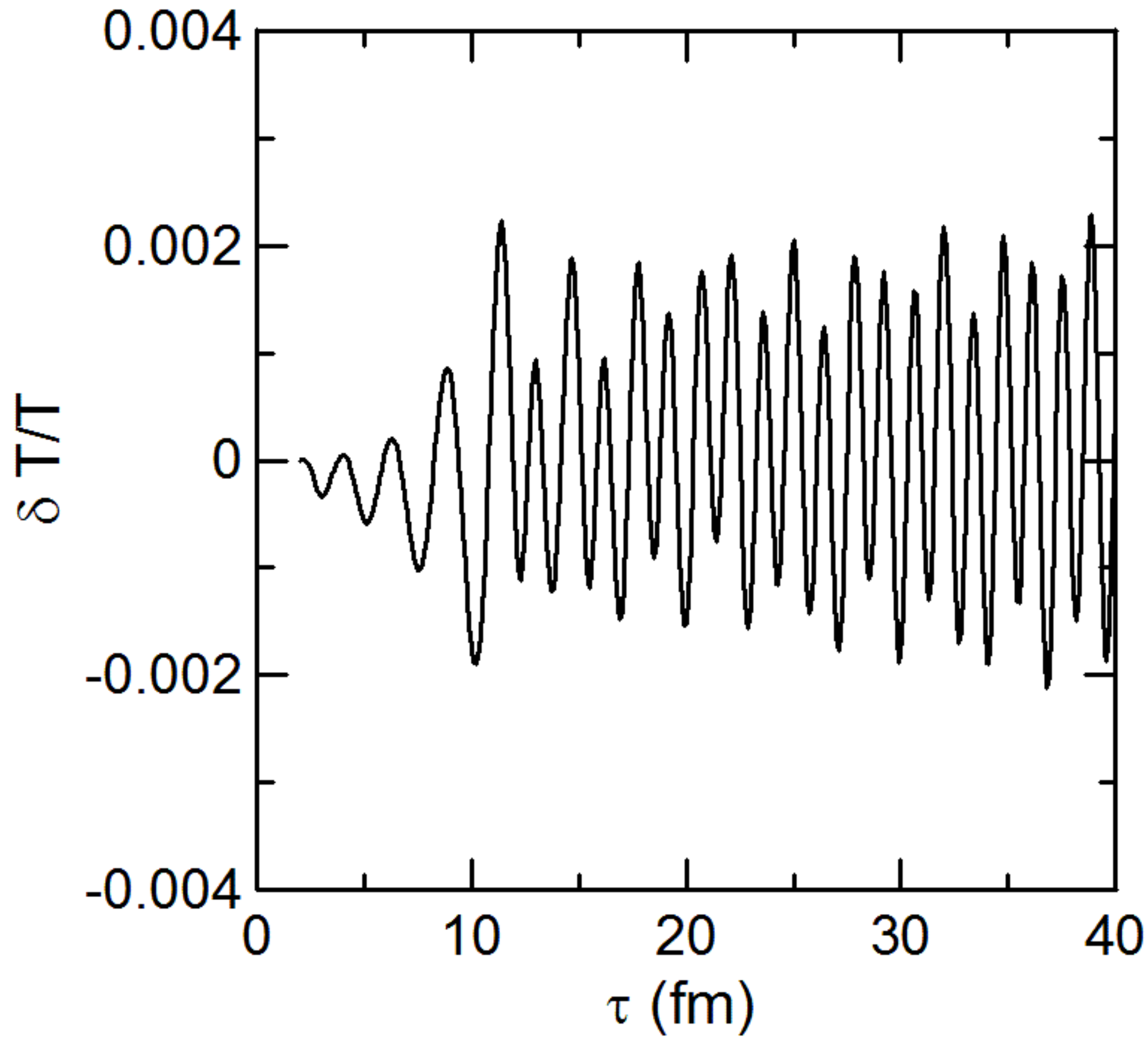}
\caption{The time dependencies of $\protect\delta \protect\sigma/\protect\sigma$ (left), $\protect\delta \protect\theta$ (middle) and $\protect\delta T/T$ (right) for $g=4.5$ at $k=1$.} \label{deltag=45}
\end{figure}

The corresponding results for $g=4.5$ and $k=1$ are shown in Fig.\ \ref{deltag=45}. One can notice a qualitative difference in the behavior of $\delta\sigma$: the amplitude of fluctuations first drops but then, at $\tau>12$ fm, it starts to grow slowly. However, this effect is rather weak, we
do not see a strong signature of the instability as observed for the static background in Sect. IV.

\subsection{Manifestation of the transient instability}

The existence of a transient instability can be explicitly demonstrated for a dynamical system as well. As follows from Eq.\ (\ref{eom-2}) in the
linear approximation, the evolution of fluctuations of the chiral field is governed by the equation
\begin{equation}
\left[ \partial _{\tau }^{2}+\frac{1}{\tau }\partial _{\tau }+\frac{k^{2}}{\tau ^{2}}-\Gamma ^{2}(\tau )\right] \delta \sigma _{k}(\tau )=-g\rho
_{T}\frac{\delta s_{0}\tau _{0}}{s_{T}\tau }~,  \label{eq_sigma_flu}
\end{equation}
where $\delta s_{0}$ is the initial fluctuation of entropy density, and other notations are explained in Eq.\ (\ref{notations}). Here we have
introduced the function
\begin{equation}
\Gamma ^{2}(\tau )\equiv -m_{\mathrm{pole}}^{2}(\tau )=-\lambda ^{2}(3\sigma ^{2}-v^{2})-g\rho _{\sigma }+g\frac{s_{\sigma }}{s_{T}}\rho _{T}~,
\label{Gamma}
\end{equation}
where $m_{\mathrm{pole}}^{2}$ is defined in Eq.\ (\ref{massgap}) and all quantities on the r.h.s. are taken from the background solution shown in Fig.\
\ref{aveboost}.

\begin{figure}[tbp]
\includegraphics[scale=0.3]{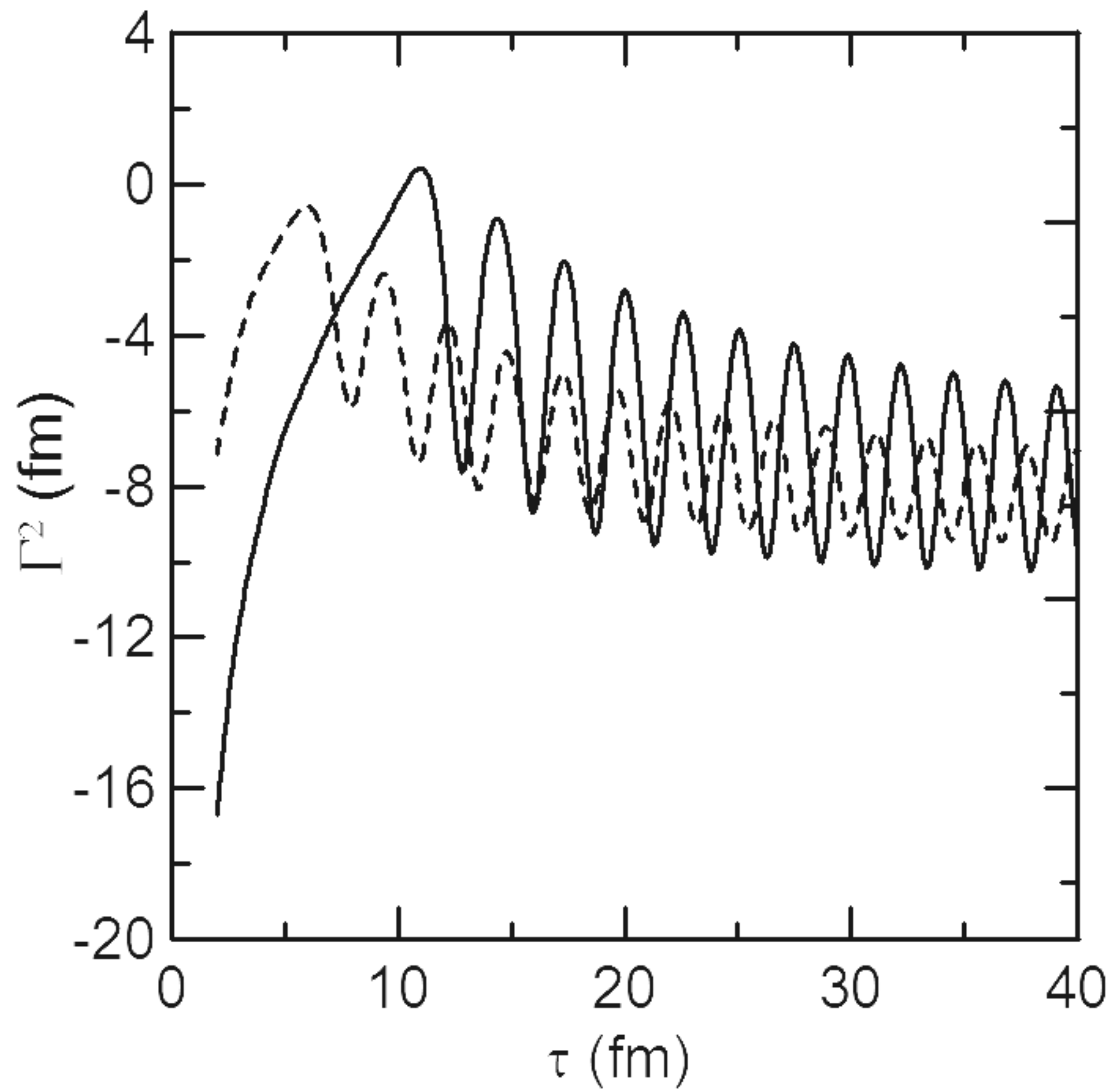}
\caption{The time dependence of $\Gamma^2_0$ for $g=3.3$ (dashed) and $4.5$ (solid). The solid line becomes negative around $\protect\tau = 12$ fm. }
\label{gammafig}
\end{figure}

The r.h.s. of Eq.\ (\ref{eq_sigma_flu}) plays the role of an external force and is irrelevant for the stability analysis. Therefore, the evolution of
fluctuations is determined by the function $\Gamma (\tau )$ only. For the static background $\Gamma $ plays the role of an increment of instability for
exponentially growing modes, $\delta \sigma \sim \exp \left( \Gamma t\right) $. In the expanding background $\Gamma (\tau )$ is itself a function of
time, and one should solve explicitly the differential equation (\ref{eq_sigma_flu}).
But some qualitative conclusions can be made by inspecting the behavior of $\Gamma ^{2}(\tau )$. By analogy with the static case, we expect that the
instability occurs when $\Gamma ^{2}(\tau )$ becomes positive. As is clear from Eq.\ (\ref{eq_sigma_flu}), the strongest instability should corresponds
to the $k=0$ mode. The time dependence of $\Gamma ^{2}(\tau )$ is shown in Fig.\ \ref{gammafig}, where the dashed and solid lines represent two cases,
$g=3.3$ and $g=4.5$, respectively. For the crossover transition ($g=3.3$) $\Gamma ^{2}$ is always negative and $\delta \sigma $ does not show any
traces of instability even at $k=0$. On the other hand, for $g=4.5$ $\Gamma ^{2}$ becomes positive around $\tau =12$ fm, i.e. exactly at the time when
the system undergoes the first order phase transition (see Fig.\ \ref{aveboost}). This analysis shows that the first order phase transition is in
fact inducing instability even for the boost-invariant background. However, the effect of the instability is weaker than what we expected from the
static background. Because of a very short time interval when $\mathrm{Re}\Gamma (\tau )>0$, this instability does not give rise to a significant
effect on the dynamics of the fluctuations. Indeed, as one can notice in Fig.\ \ref{deltag=45}, the amplitude of the fluctuations shows only a small
rise in the time interval between 12 fm/c and 40 fm/c. The effect of weakening of spinodal instability in an expanding background was earlier
demonstrated within a kinetic approach, e.g. in Ref.\ \cite{Larionov}.

It should be emphasized that within our present model there is no mechanism to dampen the fluctuations of the order parameter field except of
expansion. This is why the sigma fluctuations in Figs.\ \ref{deltag=33} and \ref{deltag=45} show persistent oscillations over a long time, as was
also observed in Ref.\  \cite{mish4}. In a more elaborate approach presented in Refs.
\cite{Nahrgang1,Nahrgang2} the field fluctuations are continuously generated
by the noise term and destroyed by the damping term. In such an approach the field oscillations will be damped out and the corresponding energy will be
transferred to the fluid, leading to its reheating. In this case one can more realistically describe the evolution of fluctuations in an expanding
system. As demonstrated in Refs. \cite{Nahrgang1,Nahrgang2}, within such an approach a strong increase of sigma fluctuations due to supercooling and reheating is
predicted in a first order phase transition, even in an expanding background. Such modifications are also needed to simulate the effect of critical
slowing down in the second order phase transition.


\section{Concluding remarks}

In this paper, we have investigated the macroscopic behaviors of quark matter using the chiral fluid dynamics. This model is derived from the linear
sigma model with constituent quarks within the framework of the mean-field approximation. In CFD, the quark degrees of freedom are mapped to the
variables of the ideal fluid. This fluid is coupled to the dynamics of the chiral order parameter $\sigma $ via the interaction term $g<\bar{q}q>\sigma
$. Two options of the model were considered which in static background lead either to the crossover or the first order chiral phase transition. We
confirmed that the equilibrium properties of CFD model are consistent with the results obtained earlier within the mean-field approximation. The
stability of the chiral fluid was investigated by analyzing the behavior of small perturbations over static and expanding backgrounds. Generally, the
excitation spectrum of the system consists of two branches, the sound branch and the sigma branch.

For the static background case, the results are rather standard. In the case of crossover transition ($g=3.3$), the uniform state is stable for small
linear perturbations at all temperatures, as was expected. On the other hand, for the first order phase transition ($g=4.5$) small perturbations around
the metastable uniform states exhibit a singular behavior characterized by the appearance of exponentially-growing modes (spinodal instability).
However, in an expanding background of the Bjorken type, the difference between these two types of phase transitions is less pronounced. In particular,
the effect of spinodal instability is significantly diminished, because of the short time a rapidly expanding system spends in the unstable region. It
should be emphasized, however, that a new important effect comes into play in a dynamical environment, namely, the supercooling to temperatures below
the lower spinodal point, point D in Fig.\ \ref{gapfig}, accompanied by strong oscillations of the chiral field. These oscillations persist for a
long time since in the present analysis we have neglected the damping terms. As demonstrated in Refs. \cite{Nahrgang2,Herold}, the model assuming
continuous generation and damping of fluctuations gives a more realistic description of the order-parameter dynamics in the course of a first order phase transition.

An important problem to study concerns the role of viscosity terms in the chiral fluid itself. Naively, viscosity effects should help dampen the
instabilities and thus make the system more stable. For small viscosities and static background this is indeed the case, see e.g. Ref.\ \cite{Pethick}.
On the other hand, as shown in Refs. \cite{Tomasik,stab1,stab2}, boost-invariant solutions of the Navier-Stokes equation becomes unstable when bulk
viscosity is large. This instability may play a significant role in the hadronization of the hot matter produced in heavy-ion collisions. Therefore, it
is worth comparing the instability-generating mechanism investigated in this work with that of Refs. \cite{Tomasik,stab2}. It is clear that they are
conceptually very different: The instabilities examined in the present work appear in the ideal hydrodynamic limit, can only be dampened by viscosity,
and should arise, to a certain extent, in \emph{any} fluid, provided there is a first-order phase transition. The dynamical instabilities examined in
Refs. \cite{Tomasik,stab2}, on the other hand, arise due to viscosity, and only in solutions where flow gradients are sufficiently large. Therefore,
one should distinguish between instabilities of the fluid and instabilities of the dynamics, i.e flow. In the future the interplay between these two
effects should be investigated within a more realistic hydrodynamic approach.

\section{Acknowledgements}

The authors thank Marlene Nahrgang, Christoph Herold, Jan Steinheimer, Joergen Randrup and Leonid Satarov for fruitful discussions. T.K. acknowledges the financial
support by CNPq. G.T. and G.D. acknowledge the financial support received from the Helmholtz International Center for FAIR within the framework of the
LOEWE program. I.M. acknowledges partial support provided by grant NSH-215.2012.2 (Russia).

\end{document}